\input amstex
\documentstyle{amsppt}
%\NoPageNumbers
\NoRunningHeads
\NoBlackBoxes
\TagsOnRight
\pageno=1
\tolerance=9999
\magnification=1200
\topmatter
\affil
Sub-Department of Quantum Statistics and Field Theory,
Department of Physics, Moscow State University,
Moscow, 119899, Russia,
e-mail: shvedov\@qs.phys.msu.su
\endaffil
\abstract
The problem of specification of  self-adjoint operators  corresponding
to singular bilinear forms is very important for applications,  such as
quantum field theory and theory of partial differential equations with
coefficient functions being distributions.  In particular,  the formal
expression $-\Delta + g\delta({\bold x})$ corresponds to a non-trivial
self-adjoint operator $\hat{H}$ in the space $L^2({\Bbb R}^d)$ only if
$d\le 3$.  For spaces of larger dimensions (this  corresponds  to  the
strongly singular  case),  the  construction of $\hat{H}$ is much more
complicated: first one should consider the space $L^2({\Bbb R}^d)$  as
a subspace of a wider Pontriagin space,  then one implicitly specifies
$\hat{H}$.  It is shown in this paper that Schrodinger,  parabolic
and  hyperbolic  equations  containing  the  operator $\hat{H}$ can be
approximated by explicitly defined systems of evolution equations of a
larger order.  The strong convergence of evolution operators taking the
initial condition of the Cauchy problem to the solution of the  Cauchy
problem is proved.
\endabstract
\keywords
strong resolvent convergence,
singular bilinear forms, Pontriagin space, Schrodinger equation,
abstract parabolic and hyperbolic equations, semigroup of operators,
self-adjoint extensions.
\endkeywords
\title
Approximations of Strongly Singular Evolution Equations
\endtitle
\author O.Yu.Shvedov
\endauthor
\email
shvedov@qs.phys.msu.su
\endemail
\endtopmatter
\document
\def\u#1#2{{\underset {#2} \to {#1} }}
\def\r#1{(#1)}

\document
math-ph/0103011

%\newpage

%\centerline
%{\bf Approximations of Strongly Singular Evolution Equations}

%\centerline{O.Yu.Shvedov}

%\baselineskip=6.1mm
%{\it
%Sub-Department of Quantum Statistics and Field Theory,
%Department of Physics, Moscow State University,
%Moscow, 119899, Russia,\\
%e-mail: shvedov\@qs.phys.msu.su
%}

\head 1. Introduction \endhead

{\bf 1}.
The main  difficulty  of  quantum  field  theory  is  the  problem  of
divergences [1] which arise since the evolution equations  of  quantum
field theory  are  ill-defined.  It  is  suitable  to investigate such
problems, making use of the  simpler  quantum  mechanical  models  which
illustrate some  of difficulties of quantum field theory.  One of such
models is the Schrodinger equation for  the  particle  moving  in  the
external singular potential:
$$
i \frac{d\psi(t)}{dt} = \hat{H}\psi(t),
\tag{1.1}
$$
where $t\in {\Bbb R}$, $\psi(t) \in {\Cal H} = L^2({\Bbb R}^d)$, while
the Hamiltonian operator $\hat{H}$ is formally written as
$\hat{H} = -\Delta + \varphi({\bold x})$,
$\varphi({\bold x})$   is   the  operator  of  multiplication  by  a
distribution.
Such models were considered in [2-14].
As an example. one can consider the function
$\varphi({\bold  x})  =  a+ g\delta({\bold x})$,  where
$a>0$, $g\in {\Bbb R}$. Then
$$
\hat{H}_g = -\Delta + a + g\delta({\bold x}).
\tag{1.2}
$$
The more  general  example  of $\hat{H}$ than \r{1.2} is the following
formal expression:
$$
\hat{H}_g \psi = \hat{T}\psi + g\chi (\chi,\psi).
\tag{1.3}
$$
Here $\hat{T}$ is a positively definite self-adjoint operator
in $\Cal H$.  Making use of the operator  $\hat{T}$, construct the
scale of Hilbert spaces
$
... \subset {\Cal H}^2 \subset {\Cal H}^1
\subset {\Cal H} = {\Cal H}^0  \subset  {\Cal  H}^{-1}  \subset  {\Cal
H}^{-2} \subset ...$.
The space ${\Cal H}^k$ is a completion of the
subspace $\cap_{n=1}^{\infty}
D(\hat{T}^n)$ of the space $\Cal H$
with respect to the norm:
$
||\psi||_k^2 = <\psi,\psi>_k = (\psi, \hat{T}^k \psi)$.
The function $\chi$ entering to eq.\r{1.3} should belong to the
space ${\Cal H}^{-k}$  for some $k$.  Expression \r{1.2} is a partial
case of  \r{1.3} for $\hat{T} = -\Delta + a$,  $\chi({\bold x}) =
\delta({\bold x}) \in {\Cal H}^{-k}$ at $k>d/2$.

To define eq.\r{1.1} mathematically, one should specify a self-adjoint
operator in ${\Cal H}$ corresponding to the formal expression
\r{1.3} (in particular, \r{1.2}).
For $\chi\in {\Cal
H}^{-2} \ {\Cal H}$ ($d\le 3$), this problem is solved as follows [2].
One should consider the restriction of the
operator $\hat{T}$ to the domain
$$
\{ \psi \in
\cap_{n=1}^{\infty}
D(\hat{T}^n) | (\hat{T}^{-k}\chi,\hat{T}^{k}\psi) =0 \}
\tag{1.4}
$$
(for the partial case \r{1.2} the domain is
$\{ \psi \in S({\Bbb R}^d)
| \psi (0) =0 \}$.  One justifies that the defect indices of
this  symmetric  operator  are (1,1).
Making use of the standard procedure (see, for example, [15]),
one constructs the one-parametric set
$\{\hat{H}_g \}$ of self-adjoint extensions of the operator
$\hat{T}$. It  is  in  one-to-one correspondence to the one-parametric
set of formal expressions \r{1.3}.

{\bf 2}. For the strongly singular case, i.e. for
$\chi  \notin   {\Cal   H}^{-2}$   ($d>3$), the operator  $\hat{T}$
considered on the domain \r{1.4} is essentially self-adjoint,  so that
the considered  approach  does not allow us to construct a non-trivial
self-adjoint operator corresponding to the formal expression
\r{1.3}.  It was noted in
[16,17,5,6] that one should consider an indefinite inner product space
instead of  the  space  $\Cal  H$  in order to construct a non-trivial
self-adjoint operator $\hat{H}$ in the strongly singular cases.

A self-adjoint operator  $\hat{H}$  in
the Pontriagin  space  $\Pi_m$  [18]  corresponding  to the expression
\r{1.3} was specified in [4] (see also [7]), provided that
$\chi  \in  {\Cal  H}^{-k-1}
\setminus {\Cal H}^{-k}$ for some $k$.  Here $m=[k/2]$.
One-parametric set of formal expressions
\r{1.3} corresponds to the
$k$-parametric set of operators $\hat{H}$  in
$\Pi_{[k/2]}(g_2,...,g_k)$;
$k-1$  parameters specifies the inner product,
while one parameter is an analog of $g$. Denote
the operator constructed in [4] (see section 2) as
$\hat{H}(g_2,...,g_k,\alpha)$.
Therefore, the equation
$$
i \frac{d\psi(t)}{dt} =\hat{H}(g_2,...,g_k,\alpha)\psi(t)
\tag{1.5}
$$
for $\psi(t) \in \Pi_{[k/2]}(g_2,...,g_k)$ is defined.

According to the analog of the Stone theorem for the Pontriagin spaces
[19,20], the operator
$\hat{H}$ is a generator of a one-parametric group of unitary operators
$e^{-i\hat{H}t}$ in $\Pi_m$. The operator
$\hat{U}(t) = e^{-i\hat{H}(g_2,...,g_k,\alpha)t}$ restricted to
$D(\hat{H}(g_2,...,g_k,\alpha))$ is an operator
taking the initial condition of the Cauchy problem
for eq.\r{1.1} to the solution of eq.\r{1.1}.

{\bf 3.}  The  problem  of  constructing  approximations  of  singular
equations \r{1.5} often arises [3]. This problem is also important for
quantum field theory [21].

It was shown in [10] that for
$m=0$ the operator transforming the initial condition for  the  Cauchy
problem to  the  solution  of the Cauchy problem for eq.\r{1.5} can be
approximated in the strong sense as $n\to\infty$
by the evolution operator for the equation
$$
i \frac{d\psi_n(t)}{dt}
=  \hat{T}  \psi_n(t)  +  g_n  \chi_n  (\chi_n,
\psi_n(t)), \qquad \psi_n(t) \in {\Cal H}
\tag{1.6}
$$
provided that
$$
||\hat{T}^{-1}\chi_n - \hat{T}^{-1} \chi|| \to_{n\to\infty} 0,
\qquad
g_n^{-1} + (\chi_n, \hat{T}^{-1} \chi_n) \to_{n\to\infty} -\alpha^{-1}.
\tag{1.7}
$$
Note that for all
$\chi\in  {\Cal  H}^{-2}$ there exist sequences
$g_n   \in   {\Bbb  R}$,  $\chi_n  \in  {\Cal  H}$
obeying \r{1.7},  for example, $\chi_n = e^{-\hat{T}/n} \chi$, $g_n =
-(\alpha^{-1} + (\chi_n, \hat{T}^{-1} \chi_n))^{-1}$.

This paper deals with the construction of an approximation for eq.\r{1.5}
for the strongly singular case (for   $m\ne   0$   or   $k>1$).
The approximation  \r{1.6}  cannot be applied then.
It happens that the resolving operator for the Cauchy problem for
eq.\r{1.5} (the $t$-dependent
operator transforming the initial condition of the
Cauchy problem  to  the  solution  of  eq.\r{1.5}  at fixed $t$)
can be viewed as a limit as $n\to\infty$  of resolving operators
for the Cauchy problem of the system of  differential  equations  of  a
larger order:
$$
\matrix
i \frac{d\psi_n(t)}{dt}  =  \hat{T}  \psi_n(t)  +
c_n(t) \chi_n, \\ z_{0,n}c_n(t) + i z_{1,n} \frac{dc_n(t)}{dt} + ... +
i^{k-1}   z_{k-1,n}   \frac{d^{k-1}   c_n(t)}{dt^{k-1}}   =   (\chi_n,
\psi_n(t)).
\endmatrix
\tag{1.8}
$$
Here $c_n(t) \in {\Bbb C}$ is a complex function,
$\psi_n(t)$ is an element of the space
$\Cal H$.
The limit should be considered in a generalized strong sense
[22,23].
The following conditions are imposed:
$$
\matrix
z_{s,n} + (\chi_n,  \hat{T}^{-s-1} \chi_n)  \to_{n\to  \infty}
g_s, \qquad s=\overline{0,k-1}, \\
||\hat{T}^{-\frac{k+1}{2}} (\chi_n-\chi)|| \to_{n\to\infty} 0.
\endmatrix
\tag{1.9}
$$
Here $g_1=-\alpha^{-1}$.

For the partial case $k=1$,  the left-hand side of the second equation
of system \r{1.8} contains only one term. Therefore, system \r{1.8} is
equivalent to   eq.\r{1.6}.  If  one  increases  $k$,  the  number  of
parameters $z_{s,n}$  is  also  increased,  so  that  the  terms  with
derivatives of  higher  orders  appear.  The  procedure of adding such
terms ("counterterms") is analogous to quantum field theory  procedure
of infinite renormalization of the wave function [1].

In particular,  the  Schrodinger  equation with the $\delta$-potential
which was constructed in [4] is formally written as
$$
i \frac{\partial\psi(x,t)}{\partial  t}  = [-\Delta + a + g \delta(x)]
\psi(x,t), \qquad x \in {\Bbb R}^d.
$$
It appears to be the limit as
$n\to\infty$ of the system
of equations on  $\psi_n(x,t)$  and $c_n(t)$
$$
\matrix
i \frac{\partial\psi_n(x,t)}{\partial  t} = [-\Delta + a]\psi_n(x,t) +
c_n(t) \chi_n(x),\\
z_{0,n}c_n(t) + ... +
i^{k-1}   z_{k-1,n}   \frac{d^{k-1}   c_n(t)}{dt^{k-1}}   =
\int dy \chi_n(y) \psi_n(y,t),
\endmatrix
$$
provided that $k=[d/2]$,
$\chi_n \to   \delta$   in the  ${\Cal  H}^{-k-1}$-norm and
sequences $z_{s,n} +  (\chi_n,  \hat{T}^{-s-1}  \chi_n)$,
$s=\overline{0,k-1}$, are convergent as  $n\to\infty$.

{\bf 4.}  Besides Schrodinger equation for the particle moving in the
singular potential, other equations appear in the applications.
Evolution of    relativistic    particle  in the external scalar field
is    described   by   the Klein-Gordon-type equation [1]
$$
\matrix
- \frac{d^2\psi(t)}{dt^2} =
\hat{H}(g_2,...,g_k,\alpha)\psi(t);
\endmatrix
\tag{1.10}
$$
The Schrodinger equation in the imaginary time is also considered
$$
\matrix
- \frac{d\psi(t)}{dt} =
\hat{H}(g_2,...,g_k,\alpha)\psi(t).
\endmatrix
\tag{1.11}
$$
After specifying the operator $\hat{H}(g_2,...,g_k,\alpha)$ eqs.\r{1.10}
and \r{1.11} become well-defined.  It happens that eq.\r{1.10} can be
approximated by the system
$$
\matrix
- \frac{d^2\psi_n(t)}{dt^2} = \hat{T} \psi_n(t) + c_n(t) \chi_n, \\
z_{0,n}c_n(t) -   z_{1,n}   \frac{d^2c_n(t)}{dt^2}  +  ...  +  (-1)^{k-1}
z_{k-1,n} \frac{d^{2k-2} c_n(t)}{dt^{2k-2}} = (\chi_n, \psi_n(t)),
\endmatrix
\tag{1.12}
$$
while the approximation for  eq.\r{1.11} is
$$
\matrix
- \frac{d\psi_n(t)}{dt} = \hat{T} \psi_n(t) + c_n(t) \chi_n, \\
z_{0,n}c_n(t) -  z_{1,n}   \frac{dc_n(t)}{dt}  +  ...  + (-1)^{k-1}
z_{k-1,n} \frac{d^{k-1} c_n(t)}{dt^{k-1}} = (\chi_n, \psi_n(t)).
\endmatrix
\tag{1.13}
$$

Therefore, the  evolution  operators  for  strongly singular evolution
equations \r{1.5},  \r{1.10},  \r{1.11} which was defined in
[4,7] with the help of complicated implicit procedure
can be approximated in the generalized strong sense
[22,23]  by evolution operators for explicitly defined systems of
equations \r{1.8}, \r{1.10}, \r{1.11}.

\head 2. Formulation of results \endhead

\subhead 2.1. Strongly singular equations \endsubhead

Remind the procedure of constructing the space
$\Pi_m$   and operator $\hat{H}$ entering to eq.\r{1.5}.

First of all, consider the space  ${\Cal P}_m$ containing
all linear combinations of the form:
$
\psi = \sum_{l=1}^{2m} c_l T^{-l} \chi + \psi_{reg}$,
where $c_l  \in {\Bbb C}$,  $\psi_{reg} \in {\Cal H}^{2m}$.  The inner
product in this space is specified by the $k-1$ real parameters
$g_2,...,g_k$. Set
$$
\matrix
(\chi, \hat{T}^{-s} \chi)_{reg} = g_s \qquad { for} \qquad s \le k,
\\
(\chi, \hat{T}^{-s} \chi)_{reg} =
(\chi, \hat{T}^{-s} \chi) \qquad { for} \qquad s \ge k+1.
\endmatrix
$$
The inner product in  ${\Cal P}_m$ is
$$
\matrix
<\psi,\psi>= \sum_{l,s=1}^{2m}    c_l^*c_s    (\chi,    \hat{T}^{-l-s}
\chi)_{reg} +
(\psi_{reg}, \psi_{reg}) +
\\
+ \sum_{s=1}^{2m} c_s (\hat{T}^m\psi_{reg}, \hat{T}^{-m-s}\chi) +
\sum_{s=1}^{2m} c^*_s (\hat{T}^{-m-s}\chi,\hat{T}^m\psi_{reg}).
\endmatrix
$$
This expression is well-defined, since
$\hat{T}^{-m-k}\chi \in {\Cal H}$ for
$k\ge 1$, while $\hat{T}^m\psi_{reg} \in {\Cal H}$.

Consider the completion [18] of the space
${\Cal  P}_m$  which is a Pontriagin space $\Pi_m$ with
$m=[k/2]$. It has the structure  $\Pi_m  =  {\Bbb
C}^{2m} \oplus {\Cal H}$:
$$
\Pi_m = \{ (\gamma,  \rho, \varphi) |
\gamma = (\gamma_1,...,\gamma_m) \in {\Bbb C}^m,
\rho = (\rho_1,...,\rho_m) \in {\Bbb C}^m,
\varphi \in {\Cal H} \}.
$$
Introduce an indefinite inner product in $\Pi_m$  as follows:
$$
<\Phi,\Phi> = \sum_{su=1}^m \gamma_s^* \gamma_u (\chi,  \hat{T}^{-s-u}
\chi)_{reg} - \sum_{s=1}^m (\gamma_s^*\rho_s + \gamma_s \rho_s^*)
+ (\varphi,\varphi).
$$
The one-to-one correspondence
$I:{\Cal P}_m \to \Pi_m$ between ${\Cal
P}_m$ and a dense subset of the space $\Pi_m$ can be specified as
$
I \{ \sum_{l=1}^{2m} c_l \hat{T}^{-l} \chi + \psi_{reg}  \}
= (\gamma, \rho, \varphi)$,
where
$$
\matrix
\gamma_1 = -c_1, ..., \gamma_m= -c_m,
\\
\rho_1 = \sum_{l=m+1}^{2m} c_l
(\chi, \hat{T}^{-l-1} \chi)_{reg} +
(\hat{T}^{-1} \chi, \psi_{reg}),\\ ...,\\
\rho_m = \sum_{l=m+1}^{2m} c_l
(\chi, \hat{T}^{-l-m} \chi)_{reg} +
(\hat{T}^{-m} \chi, \psi_{reg}),\\
\varphi = \sum_{l=m+1}^{2m} c_l\hat{T}^{-l} \chi + \psi_{reg}.
\endmatrix
$$
The following statement has been proved in  [4,7].

\proclaim{Lemma 2.1}  The continuation of the mapping $I$
is a one-to-one correspondence between the completion of the
space ${\Cal  P}_m$  and the space $\Pi_m$.
\endproclaim

Instead of the unbounded operator $\hat{H}$, it is more convenient
to define the bounded operator $\hat{H}^{-1}$.
Consider the formal equation  $\hat{H}\psi=\phi$,
$\hat{T} \psi + g\chi (\chi,\psi) = \phi$
and find (formally) $\psi$:
$
\psi =  \hat{T}^{-1}  \phi  +  \alpha  T^{-1}  \chi (\hat{T}^{-1}\chi,
\phi)$.
Here
$
\alpha = - \frac{1}{1/g+ (\chi, T^{-1}\chi)}$.
Therefore, define the operator
$\hat{H}^{-1}$ in the space ${\Cal P}_m$ as follows:
$$
\hat{H}^{-1} \phi  =  \hat{T}^{-1}  \phi  +  \alpha  \hat{T}^{-1} \chi
<\hat{T}^{-1}\chi, \phi>.
\tag{2.1}
$$
One should also specify a one-to-one correspondence between
$\alpha$ and $g$. For the case $m=0$, definition \r{2.1}
is in agreement with the approach based on self-adjoint extensions [2].

The operator \r{2.1} can be continued [4] to the space
$\Pi_m$. Thus, the operator
$\hat{H}^{-1}$  can be viewed as a continuous operator in the Pontriagin
space $\Pi_m$. It does not have zero eigenvalues for
$\alpha\ne  0$.  The inverse operator
$\hat{H}  \equiv  ({\hat H}^{-1})^{-1}$ is then  [18] a self-adjoint
(generally, unbounded)  operator in $\Pi_m$.

Therefore, space $\Pi_m$ and operator $\hat{H}$ are constructed.

\subhead 2.2. Approximation of a strongly singular equation
\endsubhead

Formulate now the main results of the paper.

The resolving operator for the Cauchy problem for system
\r{1.8} approximates the resolving operator for the Cauchy problem
for eq.\r{1.5}  in the general strong sense.  Remind the corresponding
definition [22,23].

Let ${\Cal  B}$ and ${\Cal B}_n$,  $n=1,2,...$ be Banach spaces,
$P_n: {\Cal B} \to {\Cal B}_n$, $n=1,2,...$ be a sequence of operators
with uniformly bounded norms:  $||P_n||\le  a
<\infty$ for some $n$-independent quantity  $a$.

\proclaim{Definition  2.1}  We say that a sequence of operators
$A_n:  {\Cal  B}_n \to {\Cal B}_n$,  $n=1,2,...$ is
$\{P_n\}$ - strongly convergent to operator
$A:  {\Cal B} \to {\Cal B}$,
if for all $v \in {\Cal B}$ the property
$|| P_n Av - A_nP_n v|| \to_{n\to\infty} 0$ is satisfied.
\endproclaim

Note that a generalized strong limit of a sequence of operators
depends (generally) on the choice of the sequence
$\{P_n\}$; this  fact  is  used  in the theory of the Maslov canonical
operator in abstract spaces [24,25].

\proclaim{Definition  2.2}
Let $u_n \in  {\Cal  B}_n$,  $n=1,2,...$,  $u\in  {\Cal  B}$.  We say
that a sequence $\{u_n\}$ is of the class $[u]$ (or is
$\{ P_n\}$-strongly convergent to $u$), if
$||u_n - P_nu|| \u{\rightarrow}{n\to\infty} 0$.
\endproclaim

Set ${\Cal  B}  =  \Pi_m$.  Denote by  ${\Cal  B}_n=  {\Bbb
C}^{k-1} \oplus  {\Cal  H}$  the space of sets  $\Phi_n =
(c_n^0,...,
c_n^{k-2}, \psi_n)$ of numbers $c_n^0, ..., c_n^{k-2} \in {\Bbb C}$ and
a vector $\psi \in {\Cal H}$.  Define an indefinite inner product
in the space ${\Cal  B}_n$ as follows:
$$
<\Phi_n,\Phi_n> = (\psi_n,\psi_n) + \sum_{js=0}^{k-2} c_n^{j*} c_n^s
z_{j+s+1,n}.
\tag{2.2}
$$
Here $z_{l,n}=0$ as $l\ge k$ by definition.

\proclaim{Lemma 2.2} Let $z_{k-1,n} \le 0$. Then the inner product
\r{2.2} contains $m$ negative squares.
\endproclaim

Note that the condition of lemma 2.2 is satisfied at sufficiently large $n$.

The system \r{1.8} can be presented as
a differential equation of the first order
$$
i\hat{Z}_n \frac{d}{dt} \Phi_n(t) = \hat{H}_n \Phi_n(t)
\tag{2.3}
$$
on the vector function $\Phi_n(t) \in {\Cal B}_n$. The operators $\hat{Z}_n$
and $\hat{H}_n$ are defined as
$$
\matrix
\hat{Z}_n (c_n^0,...,c_n^{k-2},\psi_n) =
(c_n^0,...,c_n^{k-3},z_{k-1,n}c_n^{k-2},\psi_n),\\
\hat{H}_n (c_n^0,...,c_n^{k-2},\psi_n) =\\
(c_n^1,...,c_n^{k-2}, (\chi_n,\psi_n) - z_{0,n}c_n^0 - ... - z_{k-2,n}
c_n^{k-2},\hat{T} \psi_n + c_n^0\chi_n).
\endmatrix
$$
Namely, after redefining
$i^l\frac{d^l}{dt^l}  c_n(t)  =
c_n^l(t)$ system \r{1.8} is taken to the form \r{2.3}.

\proclaim{Lemma 2.3}
Let $\psi_n(0) \in D(\hat{T})$,  $c_n^0(0), ...,c_n^{k-2}(0) \in {\Bbb
C}$. Then there exists a unique solution of the Cauchy problem for
eq.\r{2.3}. It continuously depends on the initial conditions
for $t\in[0,T]$.
\endproclaim

Define the operator $U_n(t): {\Bbb C}^{k-1} \oplus D(\hat{T})
\to {\Bbb C}^{k-1} \oplus D(\hat{T})$ taking the initial condition
of the Cauchy problem for eq.\r{1.9}  to the solution of the Cauchy problem.
Since the solution of the Cauchy problem continuously depends on the initial
condition, the operator $U_n(t)$ can be continued to
the space ${\Cal B}_n$. This continuation
$U_n(t): {\Cal B}_n \to {\Cal B}_n$ is unique,  provided  that  it  is
continuous.

\proclaim{Lemma 2.4}
The operator $U_n(t)$ conserves the indefinite inner product \r{2.2}.
\endproclaim

Introduce the operator $P_n:  {\Cal B}  \to  {\Cal  B}_n$  of the
form $P_n: (\gamma,
\rho,\varphi) \mapsto (c_n^0,...,c_n^{k-2},\psi_n)$ as follows.

For arbitrary $k$, set
$$
c_n^0= \gamma_1, ..., c_n^{m-1} = \gamma_m, \psi_n = -\sum_{j=0}^{m-1}
\gamma_{j+1} T^{-j-1} \chi_n + \varphi_n.
$$

For $k=2m+1$, set $\varphi_n=\varphi$.
For $k=2m$, set
$$
\matrix
\varphi_n = \varphi
+ \frac{\hat{T}^{-m}\chi_n [\rho_m - (\hat{T}^{-m}\chi_n, \varphi)]}
{(\hat{T}^{-m}\chi_n, \hat{T}^{-m}\chi_n)}
\endmatrix
$$

Specify the quantities $c_n^m$,  ...,  $c_n^{2m-1}$ from the relations:
$$
\matrix
(\hat{T}^{-1}\chi_n, \psi_n)  -  z_{m+1,n}  c_n^m  -  ...  -  z_{2m,n}
c_n^{2m-1} = \rho_1,
\\
...\\
(\hat{T}^{-m+1}\chi_n, \psi_n)  -  z_{2m-1,n}  c_n^m  -
z_{2m,n} c_n^{m+1} = \rho_{m-1},\\
(\hat{T}^{-m}\chi_n, \psi_n)    -
z_{2m,n} c_n^{m} = \rho_{m},
\endmatrix
\tag{2.4}
$$
For sufficiently  large $n$,  $c_n^m$,  ...,  $c_n^{2m-1}$ are defined
uniquely, since $z_{k-1,n}\ne 0$.
The mapping $P_n$ is constructed.

\proclaim{Lemma 2.5}
As $n\to\infty$,
$
<P_n\Phi,P_n\Phi> \to <\Phi,\Phi>$.
\endproclaim

Introduce now Hilbert inner products
in  ${\Cal  B}$ and  ${\Cal  B}_n$.

Remind that a Hilbert inner product in a   Pontriagin  space is
introduced as follows  [18].
First, an arbitrary  $m$-dimensional subspace  ${\Cal L}_m
\subset \Pi_m$  such  that  the indefinite inner product is negatively
definite on ${\Cal L}_m$, is considered.
Without loss of generality, one can consider only the case when the
subspace ${\Cal L}_m$ belongs to the domain of
$H$ [29]. Otherwise, introduce a basis
$e_i'$  in the space ${\Cal L}_m$, choose some vectors
$e_i$   form the domain of the operator  $H$ such that
the distance between $e_i$ and $e_i'$ is smaller than
${\varepsilon}$. Consider the span of the set of vectors $e_i$.
At sufficiently small ${\varepsilon}$ the inner product will be
negatively definite on the span.

By $J$ we denote the operator of the form
$J\Phi = \Phi$ at $\Phi \perp {\Cal L}_m$ and
$J\Phi = -\Phi$ at $\Phi \in {\Cal L}_m$.
According to [18], the bilinear form
$$
<\Phi,\Phi>_{{\Cal L}_m} = <\Phi, J \Phi>
\tag{2.5}
$$
specifies a positively definite Hilbert inner product.
The topologies corresponding to inner products
\r{2.5} at different ${\Cal L}_m$ are equivalent.

The inner product \r{2.5} specified the following norm
in ${\Cal B}$:
$$||\Phi|| =
\sqrt{<\Phi,\Phi>_{{\Cal L}_m}}.
\tag{2.6}
$$
To specify a norm in ${\Cal B}_n$, let us use the
following statement.
Let $\lambda$ be a sufficiently large positive number such that the
resolvent of the operator
$-\hat{Z}_n^{-1}H_n$
is defined at sufficiently large $n$.
Denote ${\Cal L}_m^n =(\hat{Z}_n^{-1} \hat{H}_n + \lambda)^{-1} P_n
(H+\lambda){\Cal L}_m$.

\proclaim{Lemma 2.6}
At sufficiently  large  $n$  the  inner  product \r{2.2} is negatively
definite on the $m$-dimensional subspace ${\Cal
L}_m^n \subset {\Cal B}_n$.
At sufficiently large  $n$   the inner product
$<\Phi,\Phi>_{P_n{\Cal L}_m}$ is positively definite on ${\Cal B}_n$
and defines a norm $||\Phi_n||  =  \sqrt{<\Phi,\Phi>_{P_n{\Cal  L}_m}}$.
\endproclaim

Lemma 2.6 implies the following lemma.

\proclaim{Lemma 2.7}
The operators
$P_n$ are uniformly bounded, $||P_n||\le a$ for some
$n$-independent constant $a$.
\endproclaim

The following  lemma  gives  necessary and sufficient condition for the
property
$\{    (c_n^0,...,c_n^{k-2},\psi_n)\in {\Cal B}_n \in
[(\gamma,\rho,\varphi)]$.   Denote   $\phi_n   =  \psi_n  +
\sum_{j=0}^{m-1} c_n^j \hat{T}^{-j-1} \chi_n$.

\proclaim{Lemma 2.8}
$\{    (c_n^0,...,c_n^{k-2},\psi_n) \} \in
[\gamma,\rho,\varphi]$ if and only if
$$
\matrix
\lim_{n\to\infty} c_n^0= \gamma_1,
...,
\lim_{n\to\infty}c_n^{m-1} = \gamma_m, \\
\lim_{n\to\infty}
||\phi_n - \varphi||=0,\\
\lim_{n\to\infty}
(\hat{T}^{-1}\chi_n, \phi_n)  -  z_{m+1,n}  c_n^m  -  ...  -  z_{2m,n}
c_n^{2m-1} = \rho_1,
\\
...\\
\lim_{n\to\infty}
(\hat{T}^{-m+1}\chi_n, \phi_n)  -  z_{2m-1,n}  c_n^m  -
z_{2m,n} c_n^{m+1} = \rho_{m-1},\\
\lim_{n\to\infty}
(\hat{T}^{-m}\chi_n, \phi_n)    -
z_{2m,n} c_n^{m} = \rho_{m}.
\endmatrix
$$
\endproclaim

In particular, lemma 2.8 shows that
the property of
$\{P_n\}$-strong convergence does not depend  on  the  choice  of  the
subspace ${\Cal L}_m$.

The following lemma shows that any initial condition for eq.
\r{1.5} can be obtained as a $\{P_n\}$-strong limit
of the sequence of initial conditions
for system \r{1.8}.

\proclaim{Lemma 2.9}
For any
$(\gamma,\rho,\varphi) \in {\Cal B}$
there exists a sequence
$\{    (c_n^0,...,c_n^{k-2},\psi_n) \in {\Cal B}_n \}$ from the class
$[\gamma,\rho,\varphi]$.
\endproclaim

To prove lemma, it is sufficient to choose
$(c_n^0,...,c_n^{k-2}, \psi_n) = P_n(\gamma,\rho,\varphi)$.

The main result of the paper is formulated as follows.

\proclaim{Theorem 1}
The sequence of operators       $U_n(t)$  is         $\{P_n\}$-strongly
convergent to $U(t)$.
\endproclaim

\proclaim{Corollary}
Let
$\{    (c_n^0(0),...,c_n^{k-2}(0),\psi_n(0)) \} \in
[(\gamma,\rho,\varphi)]$.
Then\newline
$\{    (c_n^0(t),...,c_n^{k-2}(t),\psi_n(t)) \} \in
[U^t(\gamma,\rho,\varphi)]$.
\endproclaim

For nonstrongly singular case         ($k=1$  or   $m=0$)  theorem   1
gives the result of   [10].

Formulate analogs of theorem 1 for approximations of eqs. \r{1.10}
and \r{1.11}.

\proclaim{Lemma 2.10}      Let      $\psi_n      \in      D(\hat{T})$,
$c_n^0(0),...,c_n^{k-2}(0) \in {\Bbb C}$.  Then there exists a  unique
solution of  the  Cauchy  problem for system \r{1.13}.  It continuously
depends on the initial condition.
For $\psi(0)\in D(H)$, there exists a unique  solution
of the Cauchy problem for eq.\r{1.11}.  It also continuously depends on
the initial condition.
\endproclaim

By  $\tilde{U}_n(t)$, $\tilde{U}(t)$  we denote the operators
transforming the
initial conditions for the Cauchy problems for eq.\r{1.13}, \r{1.11}
to the solution of the Cauchy
problems for eq.\r{1.13}, \r{1.11} correspondingly.

\proclaim{Theorem 2}
The sequence of operators
$\tilde{U}_n(t)$   is        $\{P_n\}$-strongly
convergent to $\tilde{U}(t)$.
\endproclaim

\proclaim{Lemma 2.11}       Let       $\psi_n(0)\in       D(\hat{T})$,
$c_n(0),...,c_n^{(2k-3)} \in {\Bbb C}$. Then there exists a unique solution
of the Cauchy problem for eq.\r{1.12}. It continuously depends on the
initial condition.
\endproclaim

Note that system \r{1.12} can be presented as
$$
- \frac{d^2}{dt^2} \Phi_n(t) = \hat{Z}_n^{-1} \hat{H}_n \Phi_n.
\tag{2.7}
$$
Introduce operators  $V_n(t)$  and  $W_n(t)$  on $D(\hat{T})$ from the
relation
$$
\Phi_n(t) = V_n(t) \Phi_n(0) + W_n(t) \frac{d\Phi_n}{dt}(0).
$$
The operator taking
$\Phi_n(0)$ to the solution of the Cauchy problem for
eq.\r{2.7} at $d\Phi_n(0)/dt=0$ is denoted as $V_n(t)$.
The operator taking $d\Phi_n(0)/dt$  to  $\Phi_n(t)$
at $\Phi_n(0)=0$ is denoted as $W_n(t)$.
Since the solution continuously depends on the initial conditions,
the operators
$V_n(t)$ and $W_n(t)$ are bounded. They are uniquely continued to
the whole space ${\Cal B}_n$.

Analogously, define the operators
$V(t)$ and $W(t)$ from the relations
$$
\psi(t) = V(t) \psi(0) + W(t) \frac{d\psi}{dt}(0),
\tag{2.8}
$$
where $\psi(t) \in D(H)$ is a solution of eq.\r{1.10}, $\psi(0)\in D(H)$,
$\dot{\psi}(0)\in D(H)$ are initial conditions.

\proclaim{Theorem 3}
The sequence of operators       $V_n(t)$  is         $\{P_n\}$-strongly
convergent to  $V(t)$.
The sequence of operators $W_n(t)$   is $\{P_n\}$-strongly
convergent to $W(t)$.
\endproclaim

\head
3. Approximation of the space and resolvent convergence
\endhead

This section deals with the proof of lemmas 2.2,
2.5-2.8. We also justify that the sequence of resolvents of
the operators $\hat{Z}_n^{-1}\hat{H}_n$ converges in a general  strong
sense to the resolvent of the operator $\hat{H}$.

{\bf 1.}
Lemma  2.2 is a corollary of the following statement. Consider the real
matrices $A$  and $B$ of the dimensions $m\times m$,  which consist of
elements $A_{ij}$ and $B_{ij}$, $i,j=\overline{1,m}$.

\proclaim{Lemma 3.1}
Let the matrix  $B$ be invertible, while the matrix  $A$ be Hermitian.
Then the quadratic form
$$
\sum_{ij=1}^m [x_i^*A_{ij} x_j + y_i^*B_{ij}x_j + x^*_iB^*_{ji} y_i]
\tag{3.1}
$$
contains $m$ negative and $m$ positive squares.
\endproclaim

\demo{Proof}
Since the matrix $A$ is Hermitian, it can be taken to the diagonal form
$U^TAU = diag[\alpha_1,...,\alpha_m]$ with  the  help  of  an  unitary
transformation. After substitution
$x_i = \sum_{s=1}^m U_{is} \xi_s$
and transformation  $\eta_s  =  \sum_{ij=1}^m B^*_{ji}
U^*_{is} y_j$ the quadratic form \r{3.1} is taken to the form
$$
\sum_{s=1}^m [\alpha_s \xi_s^*\xi_s + \xi_s^* \eta_s + \eta_s^*\xi_s]
\tag{3.2}
$$
One has
$$
\matrix
\alpha_s \xi_s^*\xi_s + \xi_s^*\eta_s + \eta_s^*\xi_s = \alpha_s
(\xi_s^* +   \alpha_s^{-1}\eta_s^*)(\xi_s   +  \alpha_s^{-1}\eta_s)  -
\alpha_s^{-1} \eta_s^*\eta_s, \qquad \alpha_s \ne 0,\\
\xi_s^* \eta_s + \eta_s^* \xi_s = \frac{1}{2}
[(\xi_s^*+\eta_s^*)(\xi_s+\eta_s) -
(\xi_s^*-\eta_s^*)(\xi_s-\eta_s) ], \qquad \alpha_s =0.
\endmatrix
$$
For both cases, the quadratic form
$\alpha_s \xi_s^*\xi_s  +  \xi_s^*\eta_s  +  \eta_s^*\xi_s$ contains
one negative and one positive  square.  Therefore,  the  form  \r{3.2}
contains $m$ positive and $m$ negative squares. Lemma 3.1 is proved.
\enddemo

\demo{Proof of lemma 2.2}
It is sufficient to justify that the quadratic form
$$
\sum_{js=0}^{k-2} c_n^{j*} c_n^s z_{j+s+1,n}
\tag{3.3}
$$
contains $m$ negative squares  (we set
$z_{l,n}=0$ for  $l\ge  k$).  Take it to the form
\r{3.1}. Consider 2 cases.

1. Let $k=2m+1$.  Denote  $x_j  =  c_n^{j-1}$,  $y_j=c_n^{m+j-1}$,
$j=\overline{1,m}$, $A_{ij}=z_{i+j-1,n}$,  $B_{ij}  =  z_{m+i+j-1,n}$,
$i,j=\overline{1,m}$. Since matrix elements $B_{ij}$ vanish
as $i+j>m+1$, while
$B_{ij}=z_{2m,n}\ne 0$ as $i+j=m+1$, $det B\ne 0$, and
the matrix  $B$  is invertible.
Therefore, the quadratic form \r{3.3}
is taken to the form \r{3.1} and contains $m$ negative squares.

2. Let    $k=2m$.  Denote     $x_j  =  c_n^{j-1}$,  $y_j=c_n^{m+j-1}$,
$j=\overline{1,m-1}$, $\sigma=c_n^{m-1}$.  The quadratic form   \r{3.3}
is taken to the form
$$
\matrix
\sum_{ij=1}^{m-1} [x_i^*\tilde{A}_{ij}x_j  +  y_j^*\tilde{B}_{ij}x_j +
x_i^*\tilde{B}_{ji}y_j] + z_{2m-1,n} \sigma^*\sigma +\\
\sum_{s=1}^{m-1}
[\sigma^* z_{m+s-1,n} x_s + \sigma z_{m+s-1,n} x_s^*]
\endmatrix
\tag{3.4}
$$
where $\tilde{A}_{ij} = z_{i+j-1,n}$,  $\tilde{B}_{ij} = z_{m+i+j-1,n}$,
$i,j=\overline{1,m-1}$. The matrix elements
$\tilde{B}_{ij}$ vanish at $i+j>m$ and are nonzero at
$i+j=m$. Therefore, the matrix $\tilde{B}$
is invertible. The formula \r{3.4} is taken to the form
$$
\matrix
\sum_{ij=1}^{m-1} [x_i^* (\tilde{A}_{ij}
- \frac{z_{i+m-1,n}z_{m+j-1,n}}{z_{2m-1,n}})
x_j  +  y_j^*\tilde{B}_{ij}x_j +
x_i^*\tilde{B}_{ji}y_j] +\\
z_{2m-1,n}
(\sigma^* + \sum_{s=1}^{m-1} \frac{z_{m+s-1,n}}{z_{2m-1,n}} x_s^*)
(\sigma + \sum_{s=1}^{m-1} \frac{z_{m+s-1,n}}{z_{2m-1,n}} x_s)
\endmatrix
\tag{3.5}
$$
Since $z_{2m-1,n}<0$, the quadratic form \r{3.5}  contains
$m$  negative squares. Lemma 2.2 is proved.
\enddemo

{\bf 2.} The following statement will be used further.

\proclaim{Lemma 3.2}
The sequence       $\Phi^{(n)}       =       (\gamma^{(n)},\rho^{(n)},
\varphi^{(n)}) \in  {\Cal  B}$  strongly converges to zero if and only if
$$
||\Phi^{(n)}||_1 =
\max_s [||\varphi^{(n)}||, |\gamma_s^{(n)}|,
|\rho_s^{(n)}|]
\tag{3.6}
$$
tends to zero.
\endproclaim

\demo{Proof}
First of all, prove the statement for the special choice of the subspace
${\Cal  L}_m$ entering to the definition of the norm \r{2.6}.
Denote by ${\Cal L}^{(0)}$  the subspace of the space  $\Cal
B$, which consists of all vectors of the form
$(\gamma,\rho,0)$. The quadratic form
$<\Phi,\Phi>$, considered on ${\Cal  L}^{(0)}$,
contains $m$ negative squares, so that for some subspace
${\Cal L}_m \subset {\Cal L}^{(0)}$ it is negatively definite.
Consider the Hilbert inner product \r{2.6} corresponding to
${\Cal L}_m$. It has the structure
$$
\sum_{sl=1}^{2m} x_s^* M_{sl} x_l + (\varphi,\varphi).
\tag{3.7}
$$
where $x_1=\gamma_1$,  ...,  $x_m=\gamma_m$,  $x_{m+1}  = \rho_1$,  ...,
$x_{2m}=\rho_m$, $M_{sl}$ is a some matrix.  Since the inner product
\r{3.7} is positively definite, the matrix $M_{sl}$ is also positively
definite.

Thus,  $\Phi^{(n)}$ strongly converges to zero if and only if
$||\varphi^{(n)}||$ tends to zero and  $||x^{(n)}||_M =
\sqrt{\sum_{sl=1}^{2m} x_s^{(n)*}    M_{sl}     x_l^{(n)}} \to 0$.
Since all norms in finite-dimensional space are equivalent, the latter
property is equivalent to $max  |x^{(n)}_s|  \to   0$.
Since all norms of the type \r{2.5} in the Pontriagin space are
equivalent, we  obtain the statement of the lemma for arbitrary choice
of ${\Cal L}_m$. Lemma is proved.
\enddemo

\proclaim{Corollary}
For some $A_1$ the following property is satisfied:
$
A_1^{-1} ||\Phi||_1 \ge ||\Phi|| \ge A_1 ||\Phi||_1$.
\endproclaim

\demo{Proof}
Suppose that statement of corollary is not satisfied.
Then it is possible to choose a sequence
$\Phi^{(n)}$ which obeys one of the following properties:
$$
\frac{||\Phi^{(n)}||_1}{||\Phi^{(n)}||} \to_{n\to\infty} 0,
\quad
\frac{||\Phi^{(n)}||}{||\Phi^{(n)}||_1} \to_{n\to\infty} 0.
$$
For definiteness, consider the first case. Consider the sequence
$
\Psi^{(n)} = \frac{\Phi^{(n)}}{\sqrt{||\Phi^{(n)}|| ||\Phi^{(n)}||_1} }
$,
tending to zero in the $||\cdot||_1$-norm and to infinity in the
$||\cdot||$-norm. This contradicts to lemma 3.2. Corollary is proved.
\enddemo

Consider the operator
$Q_n:   {\Cal  B}_n  \to  {\Cal  B}$  of the form  $Q_n:
(c_n^0,...,c_n^{k-2},\psi_n) \mapsto (\gamma_n,\rho_n,\varphi_n)$, £¤¥
$$
\matrix
\varphi_n = \psi_n + \sum_{j=0}^{m-1} c_n^j \hat{T}^{-j-1} \chi_n,\\
\gamma_n^j  = c_n^{j-1}, \qquad j=\overline{1,m},\\
\rho_n^j = (\hat{T}^{-j}\chi_n,  \varphi_n) - z_{m+j.n} c_n^m - ...  -
z_{2m,n} c_n^{2m-j}.
\endmatrix
\tag{3.8}
$$
Introduce in $\Cal B$ an additional indefinite
inner product:
$$
<\Phi,\Phi>_n =  \sum_{su=1}^m  \gamma_s^*  \gamma_u  g^{(n)}_{s+u}  -
\sum_{s=1}^m (\gamma_s^*\rho_s      -     \gamma_s     \rho_s^*)     +
(\varphi,\varphi),
\tag{3.9}
$$
where
$$
g_l^{(n)} = (\chi_n, \hat{T}^{-l} \chi_n) + z_{l-1,n}.
\tag{3.10}
$$

\proclaim{Lemma 3.3}
The following property
$<\Phi_n,\Psi_n> = <Q_n\Phi_n,  Q_n\Psi_n>_n$,
is satisfied for
$\Phi_n,\Psi_n \in {\Cal B}_n$.
\endproclaim

To prove lemma 3.3, it is
sufficient to substitute formulas \r{3.8}  to the inner product
\r{2.2}.

\proclaim{Corollary}
Let   $\Phi_n,\Psi_n \in {\Cal B}_n$ be such sequences that
$||Q_n\Phi_n|| \le C$, $||Q_n\Psi_n|| \le C$ for some       $C$. Then
$
<Q_n\Phi_n, Q_n\Psi_n> - <\Phi_n,\Psi_n> \to_{n\to\infty} 0$.
\endproclaim

\demo{Proof}
Denote    $Q_n\Phi_n = X_n = (\gamma_n,\rho_n,\varphi_n)$,
$Q_n\Psi_n = \tilde{X}_n = (\tilde{\gamma}_n,
\tilde{\rho}_n,\tilde{\varphi}_n)$. Statement of the corollary means that
$$
\sum_{su=1}^m \gamma_{n,s}^* \tilde{\gamma}_{n,u}
(g^{(n)}_{s+u} - g_{s+u}) \to_{n\to\infty} 0.
$$
This property is a corollary of lemma 3.2. Corollary is proved.
\enddemo

\proclaim{Lemma 3.4}
For some quantity $C$ that does not depend on $n$,  $\Phi$  and $\Psi$
the estimation $|<\Phi,\Psi>_n| \le C ||\Phi|| ||\Psi||$ is satisfied.
\endproclaim

\demo{Proof}
Let  $\Phi = (\gamma,\rho,\varphi)$,
$\Psi =   (\tilde{\gamma},\tilde{\rho},\tilde{\varphi})$.  It follows
from \r{3.9} that
$$
\matrix
|<\Phi,\Psi>_n| \le   \sum_{su=1}^m   (|\gamma_s|   |\tilde{\gamma}_u|
|g^{(n)}_{s+u}| + |\gamma_s| |\tilde{\rho}_s| +
|\tilde{\gamma}_s| |{\rho}_s|) + ||\varphi|| ||\tilde{\varphi}||
\le
\\
\sum_{su=1}^m ||\Phi||_1 ||\Psi||_1 (g^{(n)}_{s+u} +2) +
||\Phi||_1 ||\Psi||_1 \\
\le A_1^2 ||\Phi|| ||\Psi||
( \sum_{su=1}^m |g^{(n)}_{s+u}| + 2m^2 +1).
\endmatrix
$$
Since the sequences $g^{(n)}_{s+u}$ are convergent,
they are bounded. We obtain statement of the lemma.
\enddemo

\proclaim{Corollary}
Let $\Phi_n,  \Psi_n \in {\Cal B}$.  Then the following estimation  is
satisfied:
$
|<\Phi_n,\Psi_n>| \le C ||Q_n\Phi_n|| ||Q_n\Psi_n||$.
\endproclaim

Let us check that the sequence of the operators $Q_nP_n:  {\Cal B}
\to {\Cal   B}$   strongly  converges  to  1.  Justify  the  following
statement.

\proclaim{Lemma 3.5}   Let   $\xi\in  {\Cal  H}$  and  $||\hat{T}^{1/2}
\theta(b-\hat{T}) \xi|| \le C$  for some  $b$-independent quantity
$C$. Then  $\xi \in {\Cal H}^1 \subset {\Cal H}$.
\endproclaim

\demo{Proof}
Consider the sequence  $\xi_n   =  \theta  (n-\hat{T})  \xi$.
Suppose it to be not fundamental in  ${\Cal  H}^1$.  Then
for some ${\varepsilon}>0$  there exists such an increasing
sequence $n_1,n_2,n_3,...$ that $||\xi_{n_{2s}} -
\xi_{n_{2s-1}}||_{{\Cal H}^1}   =   ||I_{[n_{2s-1},n_{2s}]}  (\hat{T})
\xi||_{{\Cal H}^1} > {\varepsilon}$  (here $I_{[m,n]}(\lambda) =  1$
at $\lambda \in [m,n]$ and $I_{[m,n]}(\lambda) =  0$ at
$\lambda \notin [m,n]$). Therefore,
$$
(\xi, \hat{T} \theta (n_{2l}-\hat{T}) \xi) \ge
\sum_{s=1}^l (\xi,  \hat{T}  I_{[n_{2s-1},n_{2s}]}  (\hat{T}) \xi) \ge
{\varepsilon}l.
$$
For $l>   C/{\varepsilon}$,   we   obtain  a  contradiction  with  the
conditions of lemma.
Therefore, $\xi  =  \lim_{n\to\infty}  \xi_n  \in  {\Cal H}^1$.  Lemma
3.5 is proved.
\enddemo

\proclaim{Corollary}
Let   $\chi\in {\Cal H}^{-k-1}$ and $||\hat{T}^{-k/2}  \theta(b-\hat{T})
\chi|| \le C$.  Then   $\chi  \in  {\Cal  H}^{-k}$  for some
$b$-independent quantity $C$.
\endproclaim

Lemma 3.5 implies the following statement.

\proclaim{Lemma 3.6}
1. The sequence         $||\hat{T}^{-k/2}   \chi_n||$ tends to infinity as
$n\to\infty$.

2. The sequence of elements of ${\Cal       H}$  of the form
$\frac{\hat{T}^{-k/2}\chi_n}{||\hat{T}^{-k/2}\chi_n||}$         weakly
converges to zero as  $n\to\infty$.
\endproclaim

\demo{Proof}
1. Suppose that the sequence $||\hat{T}^{-k/2} \chi_n||$ does not tend
to infinity. Choose from it the bounded subsequence
$||\hat{T}^{-k/2} \chi_{n_j}|| \le C$. One has:
$$
|| \hat{T}^{-k/2}    \theta    (b-\hat{T})    \chi_{n_j}    ||     \le
||\hat{T}^{-k/2} \chi_{n_j}|| \le C.
$$
Consider the limit of the left-hand side as
$j\to\infty$. Use the fact that the operator
$\hat{T}^{1/2} \theta(b-\hat{T})$ is bounded. We obtain:
$
||\hat{T}^{-k/2} \theta(b-\hat{T}) \chi || \le C$.
It follows form lemma 3.5 and  property
$\hat{T}^{-\frac{k+1}{2}}\chi \in {\Cal H}$
that $\hat{T}^{-\frac{k}{2}}\chi \in {\Cal  H}$,  so  that
$\chi  \in {\Cal H}^{-k}$.
This contradicts to the condition
$\chi  \in {\Cal H}^{-k-1} \ {\Cal H}^{-k}$.

2. Denote $\eta_n = \frac{\hat{T}^{-k/2} \chi_n}
{||\hat{T}^{-k/2} \chi_n||}$. If $\xi\in D(\hat{T}^{1/2})$, one has:
$$
(\eta_n,\xi) = \frac{
(\hat{T}^{-\frac{k+1}{2}}\chi_n , \hat{T}^{1/2}\xi)
} {||\hat{T}^{-k/2} \chi_n||} \to_{n\to\infty} 0,
$$
since $(\hat{T}^{-\frac{k+1}{2}}\chi_n , \hat{T}^{1/2}\xi)
\to_{n\to\infty} (\hat{T}^{-\frac{k+1}{2}}\chi   ,   \hat{T}^{1/2}\xi)
\ne\infty$, $||\hat{T}^{-k/2} \chi_n|| \to_{n\to\infty} \infty$. Thus,
the sequence $\eta_n$,  $n=1,2,...$ of the elements of the unit sphere
in $\Cal  H$ weakly converges to zero on dense subset of
$\Cal H$. Therefore [22], the sequence $\eta_n$
weakly converges to zero. Lemma 3.6 is proved.
\enddemo

Lemma 3.6 implies that $z_{s,n}<0$ for sufficiently large  $n$.

\proclaim{Corollary 1}
Let $\Phi \in {\Cal B}$. The following property is satisfied:
$
Q_nP_n\Phi \to_{n\to\infty} \Phi$.
\endproclaim

\demo{Proof}
It follows from the definitions of
the operators  $Q_n$ and $P_n$ \r{3.8} and \r{2.4} that
$Q_nP_n (\gamma,\rho,\varphi)    =    (\gamma,\rho,\varphi_n)$, where
$\varphi_n=\varphi$ for odd values of $k$ and
$$
\varphi_n =    \varphi    +    \frac{\hat{T}^{-m}\chi_n    [\rho_m   -
(\hat{T}^{-m}\chi_n, \varphi)]}{(\chi_n, \hat{T}^{-2m} \chi_n)}
$$
for $k=2m$. It follows from lemma 3.5 that  $\varphi_n$
strongly converges to $\varphi$ as $n\to\infty$. Lemma 3.6 is proved.
\enddemo

\proclaim{Corollary 2}
The sequence  $Q_nP_n$ is uniformly bounded.
\endproclaim

\demo{}
Namely, any strongly convergent sequence is uniformly bounded
[26].
\enddemo

\demo{Proof of lemma 2.5}
Let $\Phi \in {\Cal B}$.
It follows from lemma 3.6 that
the sequence $||Q_nP_n\Phi||$ is  bounded.  Corollary  of  lemma  3.4
tells us that
$$
<P_n\Phi, P_n\Phi> - <Q_nP_n\Phi, Q_nP_n\Phi> \to_{n\to\infty} 0.
$$
It follows from lemma 3.6 that
$
<Q_nP_n\Phi, Q_nP_n\Phi> \to_{n\to\infty} <\Phi,\Phi>$.
We obtain statement of lemma 2.5.
\enddemo

{\bf 3.} Let us obtain the commutation rule between
operator $Q_n$  and resolvent of the operator
$\hat{Z}_n^{-1}\hat{H}_n$.

Denote by $\tilde{R}_n(\lambda)$  the operator in $\Cal
B$ that takes the set\newline
$(\gamma_{n,1},...,\gamma_{n,m},
\rho_{n,1},...,\rho_{n,m}, \varphi_n)$,
$\gamma_{n,s}, \rho_{n,s} \in
{\Bbb C}$, $\varphi_n \in {\Cal H}$,
to the set  \newline
$(\tilde{\gamma}_{n,1},...,\tilde{\gamma}_{n,m},
\tilde{\rho}_{n,1},...,\tilde{\rho}_{n,m}, \tilde{\varphi}_n)$,
which is specified from the relations
$$
\matrix
\gamma_{n,s} =  \lambda \tilde{\gamma}_{n,s} + \tilde{\gamma}_{n,s+1},
\qquad
s= \overline{1,m-1}.\\
\gamma_{n,m} = \lambda \tilde{\gamma}_{n,m} + \tilde{c}_n^m,\\
\varphi_n =   (\hat{T}+\lambda)   \tilde{\varphi}_n   +  \tilde{c}_n^m
\hat{T}^{-m} \chi_n,\\
\rho_{n,j} =   \tilde{\rho}_{n,j-1}  +  \lambda  \tilde{\rho}_{n,j}  +
g^{(n)}_{j+m} \tilde{c}_n^m, \qquad j=\overline{2,m},\\
\tilde{\rho}_{n,m} =  (\hat{T}^{-m}\chi_n,  \tilde{\varphi}_n)  -  z_{2m,n}
\tilde{c}_n^m,\\
g_1^{(n)} \tilde{\gamma}_{n,1} + ... +
g_m^{(n)} \tilde{\gamma}_{n,m}   +   g^{(n)}_{m+1}   \tilde{c}^m_n   =
\rho_{n,1} - \lambda \tilde{\rho}_{n,1},
\endmatrix
\tag{3.11}
$$
where $g_n^{(s)}$ has the form \r{3.10}.

\proclaim{Lemma 3.7}
The following property is satisfied:
$
Q_n (\hat{Z}_n^{-1} \hat{H}_n + \lambda)^{-1}  =  \tilde{R}_n(\lambda)
Q_n$.
\endproclaim

\demo{Proof}
Let   $\Phi_n = (c_n^0,...,c_n^{k-2}, \psi_n) \in {\Cal B}_n$,
$\tilde{\Phi}_n = (\hat{Z}_n^{-1}\hat{H}_n + \lambda)^{-1} \Phi_n =
(\tilde{c}_n^0,...,\tilde{c}_n^{k-2}, \tilde{\psi}_n)
\in {\Cal B}_n$.
Denote
$$
\matrix
Q_n \tilde{\Phi}_n = (\tilde{\gamma}_{n,1},...,\tilde{\gamma}_{n,m},
\tilde{\rho}_{n,1},...,\tilde{\rho}_{n,m}, \tilde{\varphi}_n),\\
Q_n {\Phi}_n = ({\gamma}_{n,1},...,{\gamma}_{n,m},
{\rho}_{n,1},...,{\rho}_{n,m}, {\varphi}_n).
\endmatrix
$$
Check that $Q_n\tilde{\Phi}_n =  \tilde{R}_n(\lambda)  Q_n\Phi_n$.  It
follows from definitions of operators
$\hat{Z}_n$ and  $\hat{H}_n$ that
$$
\matrix
c_n^0 = \lambda \tilde{c}_n^0 + \tilde{c}^1_n, \\
...,\\
c_n^{k-3} = \lambda \tilde{c}_n^{k-3} + \tilde{c}^{k-2}_n, \\
z_{k-1,n} c_n^{k-2} = \lambda z_{k-1,n} \tilde{c}_n^{k-2}  +  (\chi_n,
\tilde{\psi}_n) -    z_{0,n}   \tilde{c}_n^0   -   ...   -   z_{k-2,n}
\tilde{c}_n^{k-2},\\
\psi_n =   \lambda   \tilde{\psi}_n   +   \hat{T}   \tilde{\psi}_n   +
\tilde{c}_n^0 \chi_n.
\endmatrix
\tag{3.12}
$$
Formulas \r{3.8} imply 3 first equations of the  system  \r{3.11}.  We
obtain the 4-th and the 5-th equation form formulas for
$\rho$ and¨  $\tilde{\rho}$.
The last equation is a corollary of eqs.\r{3.12}. Lemma 3.7 is proved.
\enddemo

Denote
$$
\matrix
a_n(\lambda) =
\sum_{s=1}^{2m+1} g^{(n)}_s   (-\lambda)^{s-1-2m}   -
\lambda  (\chi_n,
\hat{T}^{-2m-1} (\hat{T}+ \lambda)^{-1} \chi_n),
\\
a(\lambda) = \lim_{n\to\infty} a_n(\lambda) =
\sum_{s=1}^{2m+1} g_s   (-\lambda)^{s-1-2m}   -
\lambda  (\chi,
\hat{T}^{-2m-1} (\hat{T}+ \lambda)^{-1} \chi).
\endmatrix
\tag{3.13}
$$

\proclaim{Lemma 3.8} Under condition $a_n(\lambda) \ne 0$,
the quantities $\tilde{\gamma}$, $\tilde{\rho}$, $\tilde{\varphi}$
are defined uniquely form the system \r{3.11}.
Under condition $a(\lambda) \ne 0$
the sequence of operators  $\tilde{R}_n(\lambda)$ being defined
for $n \ge n_0$ is strongly convergent as $n\to\infty$.
\endproclaim

\demo{Proof}
Let $(\gamma,\rho,\varphi) \in  {\Cal  B}$.  Set
$\gamma_{n}  = \gamma$,  $\rho_{n}  = \rho$,
$\varphi_{n} = \varphi$,
$\tilde{R}_n(\lambda)(\gamma,\rho,\varphi) =
(\tilde{\gamma}_n,\tilde{\rho}_n,\tilde{\varphi}_n)$.
It follows from  \r{3.11} that $\tilde{\gamma}_{n,1}$
has the form:
$$
\tilde{\gamma}_{n,1} =       (a_n(\lambda)       (-\lambda)^{2m})^{-1}
B_n(\lambda),
\tag{3.14}
$$
where
$$
\matrix
B_n(\lambda) =  - \sum_{s=1}^m g_s^{(n)} \sum_{j=0}^{s-2} (-\lambda)^j
\gamma_{n,s-j-1} + \sum_{j=0}^{m-1} (-\lambda)^j \rho_{n,j+1} +\\
(-\lambda)^m ((\hat{T}+\lambda)^{-1} \hat{T}^{-m} \chi_n, \varphi_n) -
(\sum_{j=0}^{m-1} (-\lambda)^j g^{(n)}_{m+j+1} +
\\
(-\lambda)^m
(z_{2m,n} + (\chi_n, \hat{T}^{-2m} (\hat{T}+\lambda)^{-1} \chi_n))
\sum_{j=0}^{m-1} (-\lambda)^j \gamma_{n,m-j}.
\endmatrix
$$
For $a_n(\lambda)\ne 0$, $\tilde{\gamma}_{n,1}$ is not defined.
For this case, other components of the vector
$\tilde{\gamma}_n$, vectors $\tilde{\rho}_n$ and $\tilde{\varphi}_n$
are defined uniquely from system \r{3.11}.

For $a(\lambda)\ne 0$, the sequence  $\tilde{\gamma}_{n,1}$
is convergent. We prove by induction that the sequences
$$
\matrix
\tilde{\gamma}_{n,s} = \sum_{j=0}^{s-2} (-\lambda)^j  \gamma_{s-j-1}
+ (-\lambda)^{s-1} \tilde{\gamma}_{n,1},\\
\tilde{c}^m_{n} = \sum_{j=0}^{m-1} (-\lambda)^j  \gamma_{n,m-j}
+ (-\lambda)^{m} \tilde{\gamma}_{n,1}
\endmatrix
\tag{3.15}
$$
are also convergent as $n\to\infty$. Therefore, the
sequence for elements $\Cal H$ of the form
$$
\tilde{\varphi}_n =  (\hat{T}+\lambda)^{-1}  \varphi  -  \tilde{c}_n^m
\hat{T}^{-m} (\hat{T}+\lambda)^{-1} \chi_n
\tag{3.16}
$$
is also strongly convergent as $n\to\infty$.   The sequence
$\tilde{\rho}_{n,m}$ is taken to the form
$$
\tilde{\rho}_{n,m} =   ((\hat{T}+\lambda)^{-1}   \hat{T}^{-m}  \chi_n,
\varphi) -   \tilde{c}_n^m   [z_{2m,n}   +   (\chi_n,    \hat{T}^{-2m}
(\hat{T}+\lambda)^{-1} \chi_n)]
\tag{3.17}
$$
and has a limit as $n\to\infty$. Therefore, sequences
$$
\tilde{\rho}_{n,m-s} = \sum_{j=0}^{s-1} \rho_{n,m-s-j-1}  (-\lambda)^j
+ (-\lambda)^s \tilde{\rho}_{n,m} - \sum_{j=0}^{s-1} g^{(n)}_{2m-s+j+1}
(-\lambda)^j \tilde{c}_n^m
\tag{3.18}
$$
are convergent. Therefore, the sequence
$(\tilde{\gamma}_n,\tilde{\rho}_n,\tilde{\varphi}_n)$ is convergent
in the $||\cdot||_1$-norm. Because of corollary of lemma 3.2,
it is convergent in the norm $||\cdot||$. Lemma is proved.
\enddemo

Denote
$
R(\lambda) = \lim_{n\to\infty} \tilde{R}_n(\lambda)$.
It follows from proof of lemma 3.8 that
$R(\lambda)$ is a bounded operator.

We will use further

\proclaim{Lemma 3.9}
Let $A_n:{\Cal B} \to {\Cal B}$, $n=1,2,...$ be a strongly convergent
as $n\to\infty$ sequence of operators, $A_n \to_{n \to\infty} A$
and $A_nQ_n=0$. Then $A=0$.
\endproclaim

\demo{Proof}
It follows form the condition of lemma that
$A_nQ_nP_n  =0$.  Lemma 3.6 implies that
the sequence of operators
$Q_nP_n:  {\Cal B} \to {\Cal B}$  is strongly convergent to 1,
so that  $A_nQ_nP_n  \to_{n\to\infty}  A$  in a strong sense.
Therefore, $A=0$.
\enddemo

\proclaim{Lemma 3.10}
Let  $a(\lambda) \ne 0$, $a(\mu) \ne 0$.
Then
$$
R(\lambda) - R(\mu) = (\mu-\lambda) R(\lambda) R(\mu).
\tag{3.19}
$$
\endproclaim

\demo{Proof}
Consider the following sequence of operators $A_n$:
$
A_n =  \tilde{R}_n(\lambda)   -   \tilde{R}_n(\mu)   +   (\lambda-\mu)
\tilde{R}_n(\lambda) \tilde{R}_n(\mu)$.
It satisfies the property  $A_nQ_n=0$   and strongly converges as
$n\to\infty$ to
$
R(\lambda) - R(\mu) + (\lambda-\mu) R(\lambda) R(\mu)
$.
We obtain statement of lemma.
\enddemo

\proclaim{Lemma 3.11}
Under condition $a(\lambda) \ne 0$
the following property is satisfied:
$$
R(\lambda) = (\lambda + \hat{H})^{-1}.
\tag{3.20}
$$
\endproclaim

\demo{Proof}
Justify that for $\lambda=0$ the operator  $R(\lambda)$ coincides with
the operator $\hat{H}^{-1}$ defined in section 2.
Find an explicit form of
$\hat{H}^{-1}$. It follows from \r{2.1} that
$$
\hat{H}^{-1} [\sum_{l=1}^{2m} c_l \hat{T}^{-l} \chi + \psi_{reg} ]
= \alpha a \hat{T}^{-1} \chi + \sum_{l=1}^{2m} c_l \hat{T}^{-l-1} \chi +
\hat{T}^{-1} \psi_{reg}.
$$
where $a = <\hat{T}^{-1} \chi,  \sum_{l=1}^{2m}
c_l \hat{T}^{-l} \chi + \psi_{reg}>$.

Denote
$I[\sum_{l=1}^{2m}  c_l\hat{T}^{-l}\chi  +   \psi_{reg}]   =
(\gamma,\rho,\varphi)$,
$I[\sum_{l=1}^{2m}  \alpha a \hat{T}^{-1}\chi +
c_l\hat{T}^{-l-1}\chi  +   \hat{T}^{-1} \psi_{reg}]   =
(\tilde{\gamma},\tilde{\rho},\tilde{\varphi})$.
One has:
$$
\matrix
\tilde{\varphi} =   -\gamma_m   \hat{T}^{-m-1}   \chi  +  \hat{T}^{-1}
\varphi,\\
\tilde{\gamma}_1 = - \alpha a,  \qquad  \tilde{\gamma}_2  =  \gamma_1,
\qquad,..., \tilde{\gamma}_m = \gamma_{m-1},\\
\tilde{\rho}_s = -  (\chi,  \hat{T}^{-m-s-1}  \chi)_{reg}  \gamma_m  +
\rho_{s+1}, \qquad s=\overline{1,m-1},\\
\tilde{\rho}_m =  -  (\chi,  \hat{T}^{-2m-1}  \chi)_{reg}  \gamma_m  +
(\hat{T}^{-m-1}\chi,\varphi).
\endmatrix
\tag{3.21}
$$
The formula \r{3.14} can be presented in the following form as
$n\to\infty$:
$
\tilde{\gamma}_1 = g_1^{-1} (\rho_1- \sum_{s=0}^m g_{s+1} \gamma_s)$,
For the case  $\alpha = -g_1^{-1}$ it coincides with
$\tilde{\gamma}_1 = - \alpha
a$. Formulas \r{3.15} - \r{3.18} also coincide with  \r{3.21}.
Therefore, property \r{3.20} is satisfied as
$\lambda=0$. It follows from \r{3.19} that
$R(\lambda)$ is a pseudoresolvent  [22],  Therefore, property
\r{3.20} is satisfied for all
$\lambda$ obeying the condition $a(\lambda)\ne 0$.
\enddemo

\proclaim{Lemma 3.12}
Under condition $a(\lambda)\ne 0$
the following property is satisfied:
$$
\matrix
<(\hat{Z}_n^{-1}\hat{H}_n + \lambda)^{-1} P_n \Phi,
(\hat{Z}_n^{-1}\hat{H}_n + \lambda)^{-1} P_n \Psi> \to_{n\to\infty}
\\
<(\hat{H}+\lambda)^{-1}\Phi, (\hat{H}
+\lambda)^{-1} \Psi>,  \qquad \Phi,  \Psi \in
{\Cal B}.
\endmatrix
$$
\endproclaim

\demo{Proof}
Check that the conditions of the corollary of lemma 3.3
are satisfied. Namely, for
$\Phi \in {\Cal B}$ one has
$$
\matrix
||Q_n \hat{Z}_n^{-1}\hat{H}_n   +   \lambda)^{-1}   P_n    \Phi||    =
||(\hat{H}+\lambda)^{-1} Q_nP_n   \Phi||   \le\\
||(\hat{H}+\lambda)^{-1}||  \max_n
||Q_nP_n\Phi|| \le C.
\endmatrix
$$
An analogous property is correct for
$\Psi$ also. Therefore,
$$
\matrix
<(\hat{Z}_n^{-1}\hat{H}_n + \lambda)^{-1} P_n \Phi,
(\hat{Z}_n^{-1}\hat{H}_n + \lambda)^{-1} P_n \Psi>
\\
- <(\hat{H}+\lambda)^{-1}Q_nP_n\Phi, (\hat{H}+\lambda)^{-1} Q_nP_n\Psi>
\to_{n\to\infty} 0.
\endmatrix
$$
The properties
$Q_nP_n\Phi  \to  \Phi$,  $Q_nP_n\Psi  \to  \Psi$ imply
the statement of the lemma.
\enddemo

\demo{Proof of lemma 2.6}
Choose such a  basis $e_1,...,e_m$ in ${\Cal  L}_m$ that
obeys the condition
$<e_i,e_j> =   -\delta_{ij}$.
To prove negative definiteness of the inner product on
${\Cal   L}_m^n   =
(\hat{Z}_n^{-1}\hat{H}_n + \lambda)^{-1} P_n (\hat{H}
+\lambda) {\Cal  L}_m$,  it  is  sufficient  to  check  the   positive
definiteness of the matrix
$$
A^{(n)}_{ij} = -
<(\hat{Z}_n^{-1} \hat{H}_n + \lambda)^{-1} P_n (\hat{H}+\lambda) e_i,
(\hat{Z}_n^{-1} \hat{H}_n + \lambda)^{-1} P_n (\hat{H}+\lambda) e_j>,
\tag{3.22}
$$
Its components tend to the components of the unit matrix according  to
lemma 3.12.
At sufficiently large  $n$ $||A^{(n)}-1||<1/2$, so that
$$
\matrix
(\xi,A^{(n)}\xi) -  \frac{1}{2}  (\xi,\xi)=  \frac{1}{2}  (\xi,\xi)  +
(\xi, (A^{(n)}-1)\xi)  \ge\\
\frac{1}{2}  ||\xi||^2   -   ||A^{(n)}-1||
||\xi||^2 \ge 0.
\endmatrix
$$
Positive definiteness of the inner product
$<\Phi,\Phi>_{{\Cal L}_m^n}$ is a corollary of general results of [18].
Lemma 2.6 is proved.
\enddemo

\proclaim{Lemma 3.13}
Let   $|<P_n\Phi,P_n\Phi>| \le B_1||\Phi||^2$ for some
constant $B_1$,
$
<(\hat{Z}_n^{-1}\hat{H}_n + \lambda)^{-1} P_n e_i, P_n\Phi>|
\le C_i ||\Phi||$, $i=\overline{1,m}$
for some $C_1,...,C_m$.
Then $||P_n||\le a$.
\endproclaim

\demo{Proof}
It follows from formula  \r{2.5} that:
$$
\matrix
||P_n\Phi||^2 = <P_n\Phi, P_n\Phi>_{{\Cal L}^n_m} = <P_n\Phi, P_n\Phi>
\\
+ 2 \sum_{ij=1}^m <P_n\Phi,
(\hat{Z}_n^{-1}\hat{H}_n + \lambda)^{-1} P_n e_i> M^{(n)}_{ij}
\\ \times
<(\hat{Z}_n^{-1}\hat{H}_n + \lambda)^{-1} P_n e_j, P_n\Phi>,
\endmatrix
$$
where $M^{(n)}$ is a matrix being inverse to
\r{3.22}. It follows form the conditions of lemma that
$$
||P_n\Phi||^2 \le (B_1+ 2\sum_{ij=1}^m C_i M^{(n)}_{ij} C_j)  ||\Phi||^2
\le (B_1+ 2\sup_n ||M^{(n)}|| ||C||^2) ||\Phi||^2.
$$
We obtain statement of lemma 3.13.
\enddemo

\demo{Proof of lemma 2.7}
Check that conditions of lemma 3.13 are satisfied.
Use the corollary of lemma 3.4.
$$
\matrix
|<P_n\Phi,P_n\Phi>| \le C (\sup_n ||Q_nP_n\Phi||^2 \le C_1, \\
|(\hat{Z}_n^{-1}\hat{H}_n + \lambda)^{-1} P_n e_i, P_n\Phi>| \le
\\
C \sup_n ||Q_n(\hat{Z}_n^{-1}\hat{H}_n + \lambda)^{-1} P_n e_i||
\sup_n ||Q_nP_n\Phi|| \le \\
C ||(H+\lambda)^{-1}||    (\sup_n   ||Q_nP_n||)^2   ||(H+\lambda)e_i||
||\Phi||.
\endmatrix
$$
Lemma 2.7 is proved.
\enddemo

\proclaim{Lemma 3.14}
$
||\Phi_n|| \le A_3||Q_n\Phi_n||
$
for some constant $A_3$.
\endproclaim

\demo{Proof}
One has:
$$
\matrix
||\Phi_n||^2 = <\Phi_n, \Phi_n> + \\
2 \sum_{ij=1}^m
 <\Phi_n,
(\hat{Z}_n^{-1}\hat{H}_n + \lambda)^{-1} P_n e_i> M^{(n)}_{ij}
<(\hat{Z}_n^{-1}\hat{H}_n + \lambda)^{-1} P_n e_j, \Phi_n>,
\endmatrix
$$
It follows form lemma 3.4 that
$$
\matrix
||\Phi_n||^2 \le C ||Q_n\Phi_n||^2 + \\
2  \sum_{ij=1}^m  |M^{(n)}_{ij}|
C^2 ||Q_n\Phi_n||^2       ||\tilde{R}_n(\lambda)||^2      ||Q_nP_n||^2
||(\hat{H}+\lambda)e_i|| ||(\hat{H}+\lambda)e_j||.
\endmatrix
$$
We obtain statement of lemma.
\enddemo

\proclaim{Lemma 3.15}
Let the condition $a(\lambda)\ne 0$
be satisfied.
Then the sequence of operators
$(\hat{Z}_n^{-1} \hat{H}_n  +  \lambda)^{-1}  :  {\Cal  B}_n \to {\Cal
B}_n$ is  $\{P_n\}$-strongly convergent to
the operator
$(\hat{H}+\lambda)^{-1}: {\Cal B} \to {\Cal B}$.
\endproclaim

\demo{Proof}
It is sufficient to check that for any
$\Phi \in {\Cal B}$
$$
|| (\hat{Z}_n^{-1}    \hat{H}_n    +\lambda)^{-1}    P_n\Phi   -   P_n
(\hat{H}+\lambda)^{-1} \Phi|| \to_{n\to\infty} 0.
$$
It follows from lemma 3.14 that $||\Phi_n||\to  0$, provided that
$||Q_n\Phi_n||\to 0$.
It is sufficient to prove then that
$$
||Q_n (\hat{Z}_n^{-1}    \hat{H}+\lambda)^{-1}    P_n\Phi   -   Q_nP_n
(\hat{H}+\lambda)^{-1}\Phi || \to_{n\to\infty} 0.
$$
This property is a corollary of the relation
$$
\matrix
Q_n (\hat{Z}_n^{-1}    \hat{H}_n    +\lambda)^{-1}    P_n    \Phi    =
\tilde{R}_n(\lambda) Q_nP_n\Phi \to_{n\to\infty} (H+\lambda)^{-1} \Phi.
\\
Q_nP_n(\hat{H}+\lambda)^{-1} \Phi \to_{n\to\infty} (\hat{H}
+\lambda)^{-1} \Phi.
\endmatrix
$$
Lemma 3.15 is proved.
\enddemo

\proclaim{Lemma 3.16}
For some constant
$A_2$ the estimation
$||Q_n\Phi_n|| \le A_2 ||\Phi_n||$ is satisfied.
\endproclaim

\demo{Proof}
Since the norm of the operator $J$ entering to
eq.\r{2.5} is equal to 1, the following estimation is satisfied for the
indefinite inner product:
$$
|<\Phi_n, \Psi_n>|  \le ||\Phi_n|| ||\Psi_n||,  \qquad \Psi_n,  \Phi_n
\in {\Cal B}_n.
$$
Therefore,
$$
|<\Phi_n,\Phi_n>| \le ||\Phi_n||^2, \qquad
|<\Phi_n,P_n\Phi>| \le a ||\Phi_n|| ||\Phi||.
\tag{3.23}
$$
for all $\Phi \in {\Cal B}$,  $\Phi_n \in {\Cal B}_n$.  Lemma
3.3 implies that property \r{3.23} can be presented as
$$
\matrix
|<Q_n\Phi_n,Q_n\Phi_n>_n| \le ||\Phi_n||^2,\\
|<Q_n \Phi_n, Q_nP_n \Phi>_n| \le a ||\Phi_n|| ||\Phi||.
\endmatrix
\tag{3.24}
$$
Choose $\Phi = (\tilde{\gamma},\tilde{\rho},0)$.  Then $Q_nP_n\Phi =
\Phi$. Denote $Q_n\Phi_n = (\gamma_n,  \rho_n,\varphi_n)$.  It follows
from the second property
\r{3.24} that:
$$
|\sum_{su=1}^m \gamma_{n,s}^*    \tilde{\gamma}_u    g^{(n)}_{s+u}   -
\sum_{s=1}^m (\gamma_{n,s}^*   \tilde{\rho}_s    +    \tilde{\gamma}_s
\rho_{n,s}^*) | \le a ||(\tilde{\gamma},\tilde{\rho},0)|| ||\Phi_n||.
$$
Choose $\tilde{\rho}_s^{(l)}=\delta_{sl}$,  $\tilde{\gamma}_s=0$.  For
different $l$ we obtain:
$$
|\gamma_{n,s}| \le a \max_l  ||(0,  \tilde{\rho}^{(l)},0)||  ||\Phi||_n
\le C_1 ||\Phi_n||
$$
for some constant $C_1$. Analogously,
$
|\rho_{n,s} - \sum_{u=1}^m \gamma_{n,s} g^{(n)}_{s+u} | \le
C_2 ||\Phi_n||$
for some constant $C_2$. Therefore,
$$
|\rho_{n,s}| \le C_3 ||\Phi_n||.
$$
It follows from the first inequality
\r{3.24} that:
$$
\matrix
|(\varphi_n,\varphi_n)| \le |\sum_{su=1}^m \gamma_{n,s}^* \gamma_{n,m}
g^{(n)}_{s+u} |    +
\\
|\sum_{s=1}^m    (\gamma_{n,s}^*\rho_{n,s}   +
\rho_{n,s}^* \gamma_{n,s})| + ||\Phi_n||^2 \le C_4 ||\Phi_n||^2.
\endmatrix
$$
Therefore, $||\varphi_n|| \le C_4^{1/2} ||\Phi_n||$.
For norm \r{3.6} of the vector
$Q_n\Phi_n$, the following estimation is satisfied:
$||Q_n\Phi_n||_1   \le C||\Phi_n||$.
Making use of the corollary of
lemma 3.2, we obtain statement of lemma 3.16.
\enddemo

\proclaim{Lemma 3.17}
The sequence $\{\Phi_n\}$ is of the class $[\Phi]$ if and only if
$Q_n\Phi_n \to \Phi$.
\endproclaim

\demo{Proof}
The condition $\{\Phi_n\}\in [\Phi]$ means that
$
||\Phi_n - P_n\Phi|| \to 0$.
It follows from lemmas  3.15 and  3.17 that it is equivalent to
$$
||Q_n\Phi_n - Q_nP_n\Phi|| \to 0.
\tag{3.25}
$$
Since $||Q_nP_n\Phi - \Phi||  \to  0$  according  to  lemma  3.6,  the
condition  \r{3.25} is equivalent to  $Q_n\Phi\to\Phi$. Lemma
3.17 is proved.
\enddemo

Lemma 3.17 implies lemma 2.8.

\head
4. Some properties of solutions
of evolution equations
\endhead

This section deals with investigations of
the properties of evolution operators for
eqs.\r{1.5}, \r{1.8}, \r{1.10}, \r{1.11}, \r{1.12},
\r{1.13}.
Lemmas 2.3, 2.4 and first parts of theorems 2,3 are proved.

{\bf 1.} Investigate properties  of  the  operators  entering  to  the
right-hand sides    of    evolution   equations. As usual, we   call
operators which are self-adjoint with respect to the indefinite inner
product in $\Cal  B$  or ${\Cal B}_n$ as
$J$-self-adjoint operators,  while operators being  self-adjoint  with
respect to the inner product
$<\cdot,\cdot>_{{\Cal L}_m}$ or
$<\cdot,\cdot>_{{\Cal L}^n_m}$ will be called $H$-self-adjoint.

\proclaim{Lemma 4.1}
The operators $\hat{Z}_n^{-1}     \hat{H}_n$    and  $H$  are
$J$-self-adjoint.
\endproclaim

\demo{Proof}
It follows from [18] that it is sufficient to check
that the bounded operator
$(\hat{Z}_n^{-1} \hat{H}_n+\lambda)^{-1}$
is $J$-self-adjoint for some real
$\lambda$.  Lemmas 3.3 and 3.7 imply that this property is equivalent
to selfadjointness of the operator
$\tilde{R}_n(\lambda): (\gamma,\rho,\varphi) \mapsto
(\tilde{\gamma},\tilde{\rho},\tilde{\varphi})$
with respect to the inner product
$<\cdot,\cdot>_n$. To justify the latter property, it is sufficient to
check that
for all $\Phi=(\gamma,\rho,\varphi)$ the inner product
$$
<\Phi, \tilde{R}_n(\lambda)   \Phi>_n   =   \sum_{su=1}^m   \gamma_s^*
\tilde{\gamma}_u g^{(n)}_{s+u}     -     \sum_{s=1}^m      (\gamma_s^*
\tilde{\rho}_s +     \tilde{\gamma}_s     \rho_s^*)     +    (\varphi,
\tilde{\varphi})
\tag{4.1}
$$
is real. It follows from \r{3.11} that:
$$
\matrix
\sum_{su=1}^m \gamma_s^*  \tilde{\gamma}_u  g^{(n)}_{s+u}  =   \lambda
\sum_{su=1}^m \tilde{\gamma}_s^*   \tilde{\gamma}_u   g^{(n)}_{s+u}  +
\sum_{su=1}^m g^{(n)}_{s+u-1} \tilde{\gamma}_s^* \tilde{\gamma}_u +\\
\tilde{c}_m^* \sum_{u=1}^m     g^{(n)}_{m+u}     \tilde{\gamma}_u    -
\tilde{\gamma}_1^* \sum_{u=1}^m g^{(n)}_u \tilde{\gamma}_u.\\
\sum_{s=1}^m \gamma_s^*    \tilde{\rho}_s   =   \sum_{s=1}^m   \lambda
\tilde{\gamma}_s^* \tilde{\rho}_s          +          \sum_{s=1}^{m-1}
\tilde{\gamma}_{s+1}^* \tilde{\rho}_s + \tilde{c}_m^* \tilde{\rho}_m.\\
\sum_{s=1}^m \tilde{\gamma}_s \rho_s^*
= \sum_{s=2}^m \tilde{\gamma}_s
\tilde{\rho}_{s-1}^* +     \lambda    \sum_{s=1}^m    \tilde{\gamma}_s
\tilde{\rho}_s^* + \\  \sum_{s=1}^m    g^{(n)}_{m+s}    \tilde{\gamma}_s
\tilde{c}_m^* +      \tilde{\gamma}_1      \sum_{u=1}^m      g^{(n)}_u
\tilde{\gamma}_u^*.\\
(\varphi, \tilde{\varphi})   =   (\tilde{\varphi},    (\hat{T}+\lambda)
\tilde{\varphi}) +    \tilde{c}_m^*    (\tilde{\rho}_m    +    z_{2m,n}
\tilde{c}_m^*).
\endmatrix
$$
Therefore, expression \r{4.1} is real.

Self-adjointness of the operator
$H$ is checked analogously [4,7].
Lemma 4.1 is proved.
\enddemo

Lemma  4.1 and analog of the theorem for the Pontriagin spaces
[19] imply statements of lemmas 2.3 and 2.4.

\proclaim{Lemma 4.2}
The operator $\hat{Z}_n^{-1} \hat{H}_n$ is presented as a sum
$$
\hat{Z}_n^{-1} \hat{H}_n = H_n^1 + H_n^2
\tag{4.2}
$$
of a $H$-self-adjoint operator
$\hat{H}_n^1$ and a bounded operator $\hat{H}_n^2$;
for some $n$-independent quantities $B_1$ and $B_2$
$$
\hat{H}_n^1
\ge B_1,  ||\hat{H}_n^2|| \le B_2.
\tag{4.3}
$$
The operator $\hat{H}$  is a sum
$\hat{H}_1+\hat{H}_2$
of a $H$-self-adjoint operator
$\hat{H}_1$ being semi-bounded below and
a bounded operator $\hat{H}_2$.
\endproclaim

To prove this lemma, let us prove lemmas 4.3-4.7.

\proclaim{Lemma 4.3}
The function
$
f(\lambda) = \lambda (\chi, \hat{T}^{-k} (\hat{T}+\lambda)^{-1} \chi)$
increases and tends to infinity as  $\lambda\to\infty$.
\endproclaim

\demo{Proof}
Since the operator
$\hat{T}$ is positive and self-adjoint, the difference
$$
f(\lambda_1) -   f(\lambda_2)   =   (\lambda_1-   \lambda_2)    (\chi,
\hat{T}^{-k+1} (\hat{T}+\lambda_1)^{-1} (\hat{T}+\lambda_2)^{-1}) \chi)
$$
is positive as $\lambda_1>\lambda_2$.
Thus, $f$ increases.

Check that   $f(\lambda)$ tends to infinity
as $\lambda\to\infty$. Suppose, that $f(\lambda)<C$  for some
$C$.  Then the property of positive definiteness
of the operator $\hat{T}$ implies that for all $b$
$$
\lambda (\chi, \hat{T}^{-k}\theta(b-\hat{T})
(\hat{T}+\lambda)^{-1} \chi)
\le C.
$$
Consider the limit $\lambda\to\infty$. We find:
$
(\chi, \hat{T}^{-k}\theta(b-\hat{T}) \chi)
\le C$.
According to corollary of lemma 3.5,  we obtain a  contradiction  with
the condition $\chi
\notin {\Cal H}^{-k}$. Lemma 4.3 is proved.
\enddemo

\proclaim{Lemma 4.4.}
For all $C>0$ there exist some $\lambda_0$ and $n_0$ such that for all
$\lambda >\lambda_0$ and $n > n_0$
$
f_n(\lambda) =  \lambda  (\chi_n,  \hat{T}^{-k} (\hat{T}+\lambda)^{-1}
\chi_n) > C$.
\endproclaim

\demo{Proof}
Suppose that for some $C$ for all $\lambda_0$ and
$n_0$ there exist such $\lambda>\lambda_0$ and $n > n_0$ that
$f_n(\lambda) \le C$.
Analogously to the previous subsection, we justify that the function
$f_n(\lambda)$ is increasing. This implies that
$f_n(\lambda_0) \le C$.
Therefore, for some sequence $n_p \to\infty$
$f_{n_p} (\lambda_0) \le C$. Consider a limit $p\to\infty$. We find
$f(\lambda_0) \le C$ for all $\lambda_0$. Lemma 4.4 is proved.
\enddemo

Let $\Phi  =  (\gamma,\rho,\varphi)   \in   {\Cal   B}$.   Denote
$\tilde{\Phi}_n =          \tilde{R}_n(\lambda)         \Phi         =
(\tilde{\gamma}_n(\lambda),                   \tilde{\rho}_n(\lambda),
\tilde{\varphi}_n(\lambda)) \in     {\Cal     B}$.    $\tilde{\Phi}_n$
is determined from the system
\r{3.11}.

\proclaim{Lemma 4.5}
For some constants  $\lambda_0$, $n_0$ and $A_4$
for $\lambda\ge \lambda_0$ and $n\ge
n_0$ the operator $\tilde{R}_n(\lambda)$  is  well-defined  and  obeys
properties:
$$
\matrix
|\tilde{c}_n^{m}| \le A_4 ||\Phi||_1,\\
|\tilde{c}_n^{m}| ||\lambda     \hat{T}^{-m}    (\hat{T}+\lambda)^{-1}
\chi_n|| \le A_4 ||\Phi||_1
\endmatrix
\tag{4.4}
$$
\endproclaim

\demo{Proof}
It follows form the system \r{3.11} that
$
\tilde{c}^m_n = (a_n(\lambda))^{-1} b_n(\lambda)$,
where $a_n(\lambda)$ has the form \r{3.13}, while
$$
\matrix
b_n(\lambda) = (\hat{T}^{-m}(\hat{T}+\lambda)^{-1} \chi_n,  \varphi) +
\sum_{s=1}^{m} (-\lambda)^{-s}     \rho_{m-s+1}     + \\
\sum_{s=1}^m
\sum_{l=1}^{m+1-s} g_s^{(n)} (-\lambda)^{-l-m} \gamma_{s+l-1}.
\endmatrix
$$
For some $A_5$, the following property is satisfied:
$$
|b_n(\lambda)| \le (||\hat{T}^{-m} (\hat{T}+\lambda)^{-1} \chi_n ||
+ A_5\lambda^{-1} ) ||\Phi||_1.
$$
Obtain an estimation for $a_n(\lambda)$.

1. At $k=2m$ $z_{2m,n}=0$. Lemma 4.4 implies
$$
a_n(\lambda) \ge        \frac{1}{2}       (\chi_n,       \hat{T}^{-2m}
(\hat{T}+\lambda)^{-1} \chi_n) + \frac{1}{2},
$$
for sufficiently large $\lambda_0$ and $n_0$. Therefore,
$$
|b_n(\lambda)/a_n(\lambda)| \le 2A_1 \lambda^{-1} ||\Phi||_1 -
\frac{2||\hat{T}^{-m} (\hat{T}+\lambda)^{-1} \chi_n||}
{(\chi_n, \hat{T}^{-2m} (\hat{T}+\lambda)^{-1}\chi_n)} ||\Phi||_1.
$$
The inequalities
$||\hat{T}^{-m}  (\hat{T}+\lambda)^{-1}  \chi_n  ||\le
||\hat{T}^{-m-1} \chi_n||    \le    C_1$,    $\lambda    ||\hat{T}^{-m}
(\hat{T}+\lambda)^{-1} \chi_n||^2     -     (\chi_n,     \hat{T}^{-2m}
(\hat{T}+\lambda)^{-1} \chi_n)  =  -  (\chi_n,  \hat{T}^{-2m}  \hat{T}
(\hat{T} + \lambda)^{-2} \chi_n) \le 0$ imply eq.\r{4.4}.

2. Let $k=2m+1$. Then
$$
\matrix
(-\lambda)^{-1} a_n(\lambda)     =     \sum_{s=1}^{2m+1}     g_s^{(n)}
(-\lambda)^{s-1-2m} + (\chi_n,  \hat{T}^{-2m-1} (\hat{T}+\lambda)^{-1}
\chi_n) \ge \\        \frac{1}{2}        (\chi_n,        \hat{T}^{-2m-1}
(\hat{T}+\lambda)^{-1} \chi_n) + \frac{1}{2}.
\endmatrix
$$
We obtain the following inequality:
$
|b_n(\lambda)/a_n(\lambda)| \le \lambda^{-1} C_2 ||\Phi||_1$
and eq.\r{4.4}.

Existence of the operator
$\tilde{R}_n(\lambda)$  for  $\lambda\ge  \lambda_0$  and
$n\ge n_0$ is a corollary of the proved property
$a_n(\lambda) \ne 0$. Lemma 4.5 is proved.
\enddemo

\proclaim{Lemma 4.6}
For some constant $B_3$ the following property is satisfied:
$
\lambda||\tilde{R}_n(\lambda)||_1 = \sup_{\Phi \in {\Cal B}}
\frac{\lambda ||\tilde{R}_n(\lambda)\Phi||_1}{||\Phi||_1} \le B_3$.
\endproclaim

\demo{Proof}
It follows form the second equation of the system
\r{3.11} that
$\lambda
|\tilde{\gamma}_{n,m}| \le  C_1 ||\Phi||_1$.  We obtain from the first
equation by induction that
$\lambda
|\tilde{\gamma}_{n,s}| \le C_1 ||\Phi||_1$ for $s=\overline{1,m-1}$.
It follows form the positive definiteness of the
operator $\hat{T}$ and from the third equation that
$$
\matrix
|\lambda \tilde{\varphi}_n     (\lambda)     ||     \le      ||\lambda
(\hat{T}+\lambda)^{-1} \varphi||       +      |\tilde{c}_n^m(\lambda)|
||\hat{T}^{-m} \lambda (\hat{T}+\lambda)^{-1} \chi_n|| \le\\
||\varphi||
+ A_4||\Phi||_1 \le (A_4+1) ||\Phi||_1.
\endmatrix
$$
The latter equation of the system \r{3.11} implies:
$\lambda
|\tilde{\rho}_{n,1}| \le  C_3  ||\Phi||_1$.
The 4-th equation implies
$\lambda
|\tilde{\rho}_{n,s}| \le  C_4  ||\Phi||_1$, $s=\overline{2,m}$.
We obtain statement of lemma 4.6.
\enddemo

Corollary of lemma 3.2 implies

\proclaim{Corollary}
For some constant $B_4$ the following property is satisfied:
$
\lambda||\tilde{R}_n(\lambda)||
\le B_4
$.
\endproclaim

\proclaim{Lemma 4.7}
There exist  constants
$B_5$, $\lambda_0$ and $n_0$ such that
for $\lambda\ge \lambda_0$  and  $n\ge
n_0$
$$
\matrix
\lambda ||(\lambda + \hat{Z}_n^{-1}\hat{H}_n)^{-1}|| \le B_5,\\
\lambda ||(\lambda + \hat{H})^{-1}|| \le B_5.
\endmatrix
\tag{4.5}
$$
\endproclaim

\demo{Proof}
Lemma 3.11 implies the second property of  \r{4.5}.
It follows form lemmas 3.14, 3.16 and 3.7 that
$$
\matrix
\lambda ||(\lambda + \hat{Z}_n^{-1}\hat{H}_n)^{-1}|| =
\sup_{\Phi_n\in {\Cal B}_n} \frac{
||\lambda (\lambda + \hat{Z}_n^{-1}\hat{H}_n)^{-1}\Phi_n||
}{||\Phi_n||}
\le
\\
\sup_{\Phi_n\in {\Cal B}_n} \frac{
A_3 ||Q_n\lambda (\lambda + \hat{Z}_n^{-1}\hat{H}_n)^{-1}\Phi_n||
}{A_2||Q_n\Phi_n||}
=
\sup_{\Phi_n\in {\Cal B}_n} \frac{
A_3 ||\lambda \tilde{R}_n(\lambda) Q_n\Phi_n||
}{A_2||Q_n\Phi_n||} \le B_4A_3/A_2
\endmatrix
$$
Lemma is proved.
\enddemo

\demo{Proof of lemma 4.2}
By $R_n^{\parallel}$ we denote the orthogonal with respect to the inner
product \r{2.2} projector on the subspace
${\Cal L}_m^n$,  by  $R_n^{\perp}$  denote the orthogonal projector
on $({\Cal L}_m^n)^{\perp}$.  Set
$$
\matrix
H_n^1 = R_n^{\perp} \hat{Z}_n^{-1} \hat{H}_n R_n^{\perp},\\
H_n^2 = \hat{Z}_n^{-1} \hat{H}_n - H_n^1 =
R_n^{\parallel} \hat{Z}_n^{-1} \hat{H}_n +
\hat{Z}_n^{-1} \hat{H}_n R_n^{\parallel} +
R_n^{\parallel} \hat{Z}_n^{-1} \hat{H}_n
R_n^{\parallel}.
\endmatrix
$$
Check that the operators $H_1$ and $H_2$ obey the properties \r{4.3}.
Since the inner products $<\cdot,\cdot>$ and
$<\cdot,\cdot>_{{\Cal L}_m^n}$ coincide on
$({\Cal    L}_m^n)^{\perp}$,
$H_1^n$ is a
$H$-self-adjoint operator.
Find an estimation on the norm of the operator
$H_n^2$. The operator $R_n^{\parallel}$ is rewritten as
$
R_n^{\parallel} =  -  \sum_{ij=1}^m  e_i^{(n)}   <e_j^{(n)},   \Phi_n>
M_{ij}^{(n)}$,
where $M^{(n)}_{ij}$ is a matrix being inverse to \r{3.22},
$
e_i^{(n)} = (\hat{Z}_n\hat{H}_n + \lambda)^{-1} P_n (\hat{H}
+\lambda) e_i$.
For the norm of the
operator $\hat{Z}_n^{-1} \hat{H}_n  R_n^{\parallel}$,  we  obtain  the
following estimation:
$$
||\hat{Z}_n^{-1} \hat{H}_n   R_n^{\parallel}||   \le   m^2   \max_{ij}
|M_{ij}^{(n)}| \max_i
||\hat{Z}_n^{-1}  \hat{H}_n e_i^{(n)}|| \max_i ||e_i^{(n)}||.
\tag{4.6}
$$
Since $M_{ij}^{(n)} \to_{n\to\infty} \delta_{ij}$,
$$
\matrix
\hat{Z}_n^{-1}
\hat{H}_n e_i^{(n)}  =  P_n  (H+\lambda)  e_i  -  \lambda  e_i^{(n)},\\
||e_i^{(n)}|| \le ||(\hat{Z}_n^{-1} \hat{H}_n +\lambda)^{-1}|| ||P_n||
||(H+\lambda)e||,
\endmatrix
$$
the quantity \r{4.6} is bounded uniformly with respect to
$n$. An analogous estimation
can be obtained for norms of the operators
$\hat{Z}_n^{-1}  \hat{H}_n  R_n^{\parallel}$ ¨
$R_n^{\parallel} \hat{Z}_n^{-1}  \hat{H}_n  R_n^{\parallel}$.
Therefore, $||H_n^2|| \le B$.

To check that the operator
$\hat{H}_n^1$ is  semi-bounded below,  present it as a sum of an absolutely
convergent in the norm-topology series:
$
(\lambda + \hat{H}_n^1)^{-1} = \sum_{k=0}^{\infty}
(\lambda+ \hat{Z}_n^{-1} \hat{H}_n)^{-1}
(H_n^2(\lambda+ \hat{Z}_n^{-1} \hat{H}_n)^{-1})^k$
provided that $\lambda \ge BB_5$ and  $\lambda\ge  \lambda_0$.
Namely, for this case the norm of the
$k$-th term of the series is not larger than
$\frac{B^kB_5^{k+1}}{\lambda^{k+1}}$.  Therefore, for
sufficiently large $\lambda$  and  $n\ge    n_0$   the resolvent of
the $H$-self-adjoint operator $\hat{H}_n^1$ is bounded.
Therefore, the spectrum of the operator
$\hat{H}_n^1$ is semi-bounded below by an
$n$-independent quantity. Analogously,
we prove statement of lemma 4.2 for the operator
$\hat{H}$.  Lemma 4.2 is proved.
\enddemo

Without loss of generality,  suppose that the quantity $C$ entering to
lemma 4.2  obeys the property $C>0$.  Otherwise,  one can redefine the
operators $\hat{H}_n^1$ and $\hat{H}_n^2$.

Representation \r{4.2}  and  results  of  [22]  imply  the   following
properties of evolution operators for eqs.\r{1.8}  ¨  \r{1.13}
on $[0,t]$.

\proclaim{Lemma 4.8} The following properties are satisfied:
$$
\matrix
||e^{-it\hat{Z}_n^{-1}\hat{H}_n}|| \le     e^{Bt}, \qquad
||e^{-itH}||\le
e^{Bt}.\\
||e^{-t\hat{Z}_n^{-1}H_n}|| \le e^{(B-C)t}, \qquad
||e^{-tH}|| \le e^{(B-C)t}
\endmatrix
$$
\endproclaim

\demo{Proof}
It was shown in  [22] that if $T$ is a generator for
a one-parametric semigroup $e^{-Tt}$ such that
$$
||e^{-Tt}|| \le Me^{\beta t},
\tag{4.7}
$$
while  $A$ is a bounded operator,  then  $T+A$ is also a
generator of a semigroup.
Moreover,
$
||e^{-(T+A)t}|| \le M e^{(\beta+M||A||)t}$.
The operator $i\hat{H}_n^1$ for the case of a $H$-self-adjoint
$\hat{H}_n^1$ is  a generator of a one-parametric semigroup of 
$H$-unitary operators.  This  means that property \r{4.7} is satisfied
for $M=1$, $\beta=0$. Therefore,
$
||e^{-it\hat{Z}_n^{-1}H_n}|| \le e^{||H_n^2||t} \le e^{Bt}$.
The second inequality is checked analogously.

Since the operator $\hat{H}_1$ satisfies the property
$\hat{H}_1\ge C$, it is a generator of a one-parametric semigroup,
while $||e^{-\hat{H}_n^1t}|| \le e^{-Ct}$.
We prove lemma 4.8.
\enddemo

Nota also that since $\hat{H}_n^1+\hat{H}_n^2$ is  a  generator  of  a
one-parametric semigroup, there exists a unique solution of the Cauchy
problems for eqs. \r{1.13} and \r{1.11}  for
$\Phi_n(0)\in D(\hat{H}_n^1+\hat{H}_n^2)$.  This  solution continuously
depends on the initial conditions. Lemma 2.10 is proved.

To prove lemma 2.11,  justify some auxiliary statements being analogous
to [27].

Consider the following differential equation in the Banach space
$\Cal   B$
$$
-\frac{d^2\Phi}{dt^2} = \hat{A}\Phi,
\qquad \Phi(t) \in D(\hat{A}) \subset {\Cal B},
\qquad t\in [0,T].
\tag{4.8}
$$
with closed operator $\hat{A}$.

\proclaim{Definition 4.1}
We say that the Cauchy problem for
eq.\r{4.8} is formulated uniformly correct if for all
$\Phi(0)$ and $\dot{\Phi}(0)$ from $D(\hat{A})$
there exists a unique two times continuously differentiable
function $\Phi(t)   \in   D(\hat{A})$
satisfying eq.\r{4.8}  and initial conditions.
The dependence of $\Phi(t)$ from initial conditions
is uniformly continuous.
\endproclaim

Define on $D(A)$ the operators $V(t)$ and  $W(t)$  from  the  property
\r{2.8},
$
\Phi(t) = V(t) \Phi(0) + W(t) \dot{\Phi}(0)$.
Denote by $\dot{V}(t)$ and  $\dot{W}(t)$ the operators
from  $D(A)$  to
$\Cal B$ which are defined from the relation
$
\dot{\Phi}(t) = \dot{V}(t) \Phi(0) + \dot{W}(t) \dot{\Phi}(0)$.

Let $\Cal   B$  be a Hilbert space.

\proclaim{Lemma 4.9}
Let  $A$ be a  $H$-self-adjoint semi-bounded below operator in $\Cal
B$: $A \ge C_1 >0$.  Then the Cauchy problem for
eq.\r{4.8} is uniformly correct and
$$
||V(t)|| \le 1, \qquad ||W(t)|| \le 1/\sqrt{C_1}.
\tag{4.9}
$$
\endproclaim

\demo{Proof}
The function of the form
$$
\Phi(t) =  \cos(\sqrt{\hat{A}}t)
\Phi(0) + \frac{\sin(\sqrt{\hat{A}}t)}{\sqrt{\hat{A}}}
\dot{\Phi}(0)
\tag{4.10}
$$
is a solution of the Cauchy problem for eq.\r{4.8} [27].  Prove
the property of uniqueness. Let
$\Phi(0)=0$,   $\dot{\Phi}(0)=0$.  Consider the
function
$
f(t) =   \frac{1}{2}   (\dot{\Phi}(t),  \dot{\Phi}(t))  +  \frac{1}{2}
(\Phi(t), \hat{A} \Phi(t))$.
It satisfies the conditions   $f(0)=0$,   $df/dt=0$. Therefore,
$f(t)=0$. Since the operator $A$ is semi-bounded below,
one has
$(\dot{\Phi},\dot{\Phi})=0$, $(\Phi,\hat{A}\Phi)=0$.  Therefore, $\Phi=0$.
The property of uniqueness is proved. It follows from the explicit
form of solution of eq.\r{4.10} the property of uniform correctness of
the Cauchy problem and relations
\r{4.9}.  Lemma 4.10 is proved.
\enddemo

Suppose that there exists such
$\zeta$ that the operator
$(A+\zeta)^{-1}$ is well-defined.

\proclaim{Lemma 4.10}  Let the Cauchy problem
for eq.\r{4.8} be uniformly correct. Consider the equation
$$
-\frac{d^2\Phi(t)}{dt^2} = \hat{A}\Phi(t) + \xi(t),
\qquad \Phi(t) \in D(\hat{A}) \subset {\Cal B},
\qquad t\in [0,T].
\tag{4.11}
$$
where $\xi(t)  \in  D(\hat{A}^2)$.
$(\hat{A}+\zeta)^2\xi(t)$  is a continuous function on
$[0,T]$. Then the Cauchy problem for eq.\r{4.11} has a unique solution
of the form:
$$
\Phi(t) = V(t) \Phi(0) + W(t) \dot{\Phi}(0) - \int_0^t d\tau W(t-\tau)
\xi(\tau).
\tag{4.12}
$$
\endproclaim

\demo{Proof}
The uniqueness is obvious. Let
$\Phi_1$  and  $\Phi_2$  be two solutions of the Cauchy problem.
Then their difference satisfies eq.\r{4.8} and zero initial condition.
It follows   from  uniform  correctness  of  the  Cauchy  problem  for
eq.\r{4.8} that $\Phi_1-\Phi_2=0$.

To prove lemma, it is sufficient to justify that the function
$$
\Phi(t) = - \int_0^t d\tau W(t-\tau) \xi(\tau)
$$
obeys eq,\r{4.11} and zero initial condition. Check that
$$
\frac{d\Phi(t)}{dt} = - \int_0^t d\tau \dot{W}(t-\tau) \xi(\tau).
\tag{4.13}
$$
Consider the difference
$$
\matrix
-\frac{\Phi(t+\delta t) - \Phi(t)}{\delta t} + \dot{\Phi}(t) =\\
\int_t^{t+\delta t}  \frac{d\tau}{\delta  t}   W(t+\delta   t   -\tau)
\xi(\tau) +    \int_0^t    d\tau    (\frac{W(t+\delta   t   -\tau)   -
W(t-\tau)}{\delta t} -  \dot{W}(t-\tau)) \xi(\tau) =\\
\int_0^1 ds  W(\delta  t(1-s))  \xi(t+\delta  t  s)  +  \int_0^t d\tau
\int_0^1 ds (\dot{W}(t+ \delta t s -\tau) - \dot{W}(t-\tau)) \xi(\tau)
= \\
\int_0^1 ds W(\delta t (1-s)) \xi(t) + \int_0^1 ds W(\delta  t  (1-s))
(\xi(t+s\delta t)  - \xi(t)) +\\
\delta t
\int_0^t d\tau \int_0^1 ds \int_0^s ds'
AW(t+s'\delta t-\tau) \xi(\tau).
\endmatrix
$$
The norm of this expression is not larger than
$$
\matrix
\int_0^1 ds ||W(\delta t(1-s))|| (||\xi(t)||  +  ||\xi(t+s\delta  t)  -
\xi(t)|| +
\\
\delta t
\int_0^t d\tau \int_0^1 ds \int_0^s ds'
||\hat{A} W(t+s'\delta t-\tau) \xi(\tau)||.
\endmatrix
$$
According to the Lesbegue theorem  (see, for example, [28])
this expression tends to zero as  $\delta  t\to  0$.
Therefore, property  \r{4.13} is checked.
Initial conditions are obviously satisfied. Check
eq.\r{4.11}. One has:
$$
\matrix
-\frac{\dot{\Phi}(t+\delta t)  -  \dot{\Phi}(t)}{\delta t} - A\Phi(t) -
\xi(t) = \int_0^1 ds (\dot{W}(\delta t(1-s)) \xi(t+s\delta t) - \xi(t))
\\
+ \int_0^1   d\tau  \int_0^1  ds  (- W(t-\tau+s\delta  t)  +  W(t-\tau))
\hat{A}\xi(\tau).
\endmatrix
$$
According to the Lesbegue theorem, this expression tends to zero.
Lemma is proved.
\enddemo

\proclaim{Corollary}
Let the function $\xi(t)\in {\Cal B}$ is continuous on  $[0,T]$,  while
the function $\Phi(t)$ is a solution of eq.\r{4.11}.  Then
formula \r{4.12} is satisfied.
\endproclaim

\demo{Proof}
It is sufficient to consider the case if initial conditions vanish:
the general case can be reduced to it by the substitution
of  $\Phi(t)$ by  $\Phi(t)  -
V(t)\Phi(0) -   W(t)\dot{\Phi}(0)$.   Consider the
function  $v(t)   =
(\hat{A}+\zeta)^{-2} \Phi(t)$ satisfying the following equation:
$$
-\frac{d^2v(t)}{dt^2} = \hat{A} v(t) + (\hat{A}+\zeta)^{-2}\xi(t),
$$
and zero initial condition. Therefore,
$$
v(t) = - (\hat{A}+\zeta)^{-2} \int_0^t d\tau W(t-\tau) \xi(\tau).
$$
We obtain statement of corollary.
\enddemo

It happens that the condition that
$(\hat{A}+\zeta)^{-2}\xi$ is  continuous  can  be  substituted  by  the
condition that $\xi$ is two times continuously differentiable.

\proclaim{Lemma 4.11.}
Let all the conditions of lemma 4.10 be satisfied, except for
continuity of $(\hat{A}+\zeta)^{-2}\xi$. Let also
the function   $\xi(t)$  be two times continuously differentiable and
$\xi(0)\in D(\hat{A})$. Then statement of lemma 4.10 is satisfied.
\endproclaim

\demo{Proof}
The property of uniqueness of the solution of the Cauchy problem
is checked analogously to lemma 4.10. Corollary of lemma 4.10 tells us
that the solution of the Cauchy problem is given by formula  \r{4.12},
provided that it exists. It is sufficient then to check that expression
\r{4.12} satisfies eq.\r{4.11} and initial condition.
It is sufficient to consider the
case $\Phi(0)=0$, $\dot{\Phi}(0)=0$. Denote
$W_1(t) =  \int_0^t  d\tau  W(\tau)$,
$W_2(\tau) = \int_0^{\tau}
W_1(\tau)$.
Substituting $\xi(t) = \xi(0) + \int_0^t d\tau \xi(\tau)$, we find:
$$
\int_0^t d\tau
W(t-\tau) \xi(\tau) = W_1(\tau) \xi(0) + \int_0^t ds W_1(t-s)
\dot{\xi}(s).
$$
Applying this formula again, we obtain:
$$
\matrix
\int_0^t W(t-\tau) \xi(\tau) d\tau = \\
W_1(t)\xi(0) + W_2(t)\dot{\xi}(0)
+ \int_0^t ds W_2(t-s) \ddot{\xi}(s).
\endmatrix
\tag{4.14}
$$
It follows from the definition of the
operator $W$ that it satisfies the following equation:
$$
\ddot{W}(t) \Phi = - \hat{A}W(t) \Phi, \qquad \Phi \in D(A)
\tag{4.15}
$$
and commutes on  $D(A)$ with the operator  $A$.  Integrating twice
eq.\r{4.15}, we find:
$$
-\hat{A}W_2(t) = W(t) - W(0) - \dot{W}(0)t = W(t) - t
\tag{4.16}
$$
on $D(A)$.  The operator \r{4.16} is bounded
and can be therefore continued on $\Cal B$.
It follows from eqs.\r{4.16} and \r{4.14} that
$$
\matrix
-\hat{A}\int_0^t W(t-\tau)  \xi(\tau)  d\tau  =\\
\int_0^t  ds [W(t-s)-(t-s)]
\ddot{\xi}(s) + \dot{W}(t) \xi(0) + (W(t) - t) \dot{\xi}(0).
\endmatrix
\tag{4.17}
$$
Furthermore,
$$
\frac{d^2}{dt^2}[\int_0^t W(\tau) \xi(t-\tau) d\tau]
= W(t)\dot{\xi}(0) + \dot{W}(t)\xi(0) +
\int_0^t W(\tau) \ddot{\xi}(t-\tau) d\tau.
\tag{4.18}
$$
Comparing eqs.\r{4.17} and \r{4.18}, we obtain statement of lemma.
\enddemo

\proclaim{Lemma 4.12}
Let the operator $\hat{A}$ be a sum of a $H$-self-adjoint semi-bounded
below operator $\hat{T}_1 \ge  C_1  >0$  and a bounded operator
$\hat{T}_2$,  $||\hat{T}_2||  \le C_2$.  Then the Cauchy problem
for eq.\r{4.8} is uniformly correct and
$$
||V(t)|| \le e^{C_2t/\sqrt{C_1}},\qquad
||W(t)|| \le e^{C_2t/\sqrt{C_1}}/\sqrt{C_1}.
$$
\endproclaim

\demo{Proof}
According to corollary of lemma 4.10, the function
$\Phi(t)$ is a solution of the Cauchy problem for eq.
\r{4.8} if and only if
$$
\Phi(t) =  V_1(t)  \Phi(0)  +  W_1(t)  \dot{\Phi}(0)  - \int_0^t d\tau
W_1(t-\tau) \hat{T}_2 \Phi(\tau),
\tag{4.19}
$$
where
$V_1(t) = \cos(\sqrt{\hat{T}_1}t)$,
$W_1(t) = \frac{\sin(\sqrt{\hat{T}_1}t)}{\sqrt{\hat{T}_1}}$.
The abstract Volterra equation \r{4.19} has a unique solution
(see, for example,  proof of [28]), which can be presented as a sum of
an absolutely convergent in the norm-topology series:
$$
\matrix
\Phi(t) = \sum_{n=0}^{\infty} (-1)^n \int_{t.\tau_1> ,,, > \tau_n >0 }
d\tau_1 ...  d\tau_n  W_1(t-\tau_1)
\hat{T}_2 ...  W_1(\tau_{n-1} - \tau_n)
\hat{T}_2 \\ \times
(V_1(\tau_n) \Phi(0) + W_1(\tau_n) \dot{\Phi}(0)).
\endmatrix
$$
Therefore,
$$
\matrix
||V(t)|| \le \sum_{n=0}^{\infty} \frac{(C_2t/\sqrt{C_1})^n}{n!} =
e^{C_2t/\sqrt{C_1}}, \\
||V(t)|| \le                                       \sum_{n=0}^{\infty}
\frac{(C_2t/\sqrt{C_1})^n}{n!\sqrt{C_1}} =
e^{C_2t/\sqrt{C_1}}/\sqrt{C_1}.
\endmatrix
$$
Lemma 4.12 is proved.
\enddemo

Lemmas 4.12 and 4.2 imply

\proclaim{Corollary}
The statement of lemma 2.11 is satisfied.
For  $t\in  [0,T]$ there exists an
$n$-independent quantity $M$ such that
$||V_n(t)|| \le M$, $||V(t)|| \le M$, $||W_n(t)|| \le M$,
$||W(t)|| \le M$.
\endproclaim

\head
5. Convergence in generalized strong sense.
\endhead

Let us justify the property of generalized strong convergence
of the operators  $U_n$,  $V_n$  and
$W_n$ entering   to  theorems  1-3.  Let  us  first  investigate  some
properties of generalized strong convergence.
Formulate an analog of the Banach-Steinhaus theorem.

\proclaim{Lemma 5.1.}
Let $A_n:  {\Cal B} \to {\Cal B}_n$,  $n=1,2,...$  be  a  sequence  of
operators satisfying the property
$||A_n||  \le  M  <\infty$  for some
$n$-independent constant $M$;  ${\Cal D} \subset {\Cal
B}$ - is a dense subset of $\Cal B$, $||A_nv||\to_{n\to\infty} 0$
for $v\in  {\Cal  D}$.  Then  $||A_nv||\to_{n\to\infty}  0$ for $v\in
{\Cal B}$.
\endproclaim

\demo{Proof}
Let $v\in {\Cal B}$,  ${\varepsilon}>0$.  Choose such $v'\in {\Cal
D}$ that $||v-v'|| \le \frac{{\varepsilon}}{2M}$. Choose such $n_0$,
that for $n\ge n_0$ $||A_nv'||\le {\varepsilon}/2$. Then $||A_nv|| \le
||A_nv'|| + ||A_n|| ||v-v'|| \le {\varepsilon}$.  We obtain  statement
of lemma.
\enddemo

Remarks. The proof of ref.[26] of the Banach-Steinhaus theorem
cannot be generalized to the case of
$\{P_n\}$   -  strong convergence.
Proof of  [23]   uses also the condition
$||P_nv|| \to_{n\to\infty} ||v||$.

\proclaim{Lemma 5.2}
Let $A_n:{\Cal  B}  \to  {\Cal  B}_n$,  $n=1,2,...$  be  a sequence of
operators satisfying the following property:
for each  $v\in  {\Cal B}$
the sequence $||A_nv||$ is bounded.
Then   $||A_n||\le  M$ for
some $n$-independent quantity $M$.
\endproclaim

\demo{Proof}
is analogous to [26].
\enddemo

\proclaim{Lemma 5.3.}
Let $B_n:{\Cal    B}_n    \to    {\Cal    B}_n$,    $n=1,2,...$  be
a sequence of operators which
$\{P_n\}$-strongly converges to the operator
$B:   {\Cal   B}\to{\Cal   B}$.  Then the sequence
$||B_nP_n||$ is bounded.
\endproclaim

\demo{Proof}
Denote $A_n=B_nP_n$.  For all  $v\in  {\Cal  B}$  $||A_nv-P_nBv||
\to_{n\to\infty} 0$,  so that the sequence $||A_nv  -  P_nBv||$
is bounded, $||A_nv    -   P_nBv||\le   M$.   Therefore,
$||A_nv||   \le
||A_nv-P_nBv|| + ||P_n|| ||Bv|| \le M+a ||Av||$. Lemma 5.2 implies
statement of lemma.
\enddemo

\proclaim{Lemma 5.4}
Let $u_n\in {\Cal B}_n$, $n=1,2,...$ is a sequence of
vectors from the class
$[u]$,  $u\in {\Cal B}$,  $A_n:  {\Cal B}_n  \to  {\Cal  B}_n$,
$n=1,2,...$ is a uniformly bounded  ($||A_n|| \le M$)
sequence of operators which $\{P_n\}$-strongly converges to
the operator
$A: {\Cal  B}  \to  {\Cal  B}$.
Then the sequence $\{A_nu_n\}$
is of the class $[Au]$.
\endproclaim

\demo{Proof}
One has:
$$
||A_nu_n -  P_nAu  ||  \le  ||A_n|| ||u_n-P_nu|| + ||A_nP_nu - P_nAu||
\to_{n\to\infty} 0.
$$
\enddemo

Proofs of theorems 1 and 2 are identical to ref.[22].

\demo{Proof of theorem 3}
Let $v\in   {\Cal   B}$,   $\zeta$  satisfy the condition
$a(\lambda)\ne 0$.
Consider the function $w_n(t)$ of the form
$$
v_n(t) =   V_n(t)   (\hat{Z}_n^{-1}   \hat{H}_n    +\zeta)^{-1}    P_n
(\hat{H}+\zeta)^{-1} v  -  (\hat{Z}_n^{-1}  \hat{H}_n  +\zeta)^{-1} P_n V(t)
(\hat{H} +\zeta)^{-1}v.
$$
It obeys the following condition
$$
- \frac{d^2v_n(t)}{dt^2} = \hat{Z}_n^{-1} \hat{H}_n w_n(t) + \xi_n(t).
\tag{5.1}
$$
where
$$
\xi_n(t) =    (P_n(\hat{H}+\zeta)^{-1}
-   (\hat{Z}_n^{-1}   \hat{H}_n   +
\zeta)^{-1} P_n) V(t) v.
$$
The initial condition for eq.\r{5.1} has the form
$v_n(0)=0$, $\dot{v}_n(0)=0$.
Corollary of lemma 4.10 implies that
$$
v_n(t) = \int_0^t d\tau W_n(t-\tau) \xi_n(\tau).
$$
Therefore,
$$
||w_n(t)|| \le M \int_0^t d\tau ||\xi_n(\tau)||.
\tag{5.2}
$$
For each  $\tau$   $||\xi_n(t)||$ tends to zero
because of lemma 3.15.
Furthermore,
$$
||\xi_n(\tau)|| \le (||(\hat{Z}_n^{-1}\hat{H}_n + \zeta)^{-1} P_n||  +
||P_n|| ||(\hat{H}+\zeta)^{-1}||) M ||v||,
$$
so that  the sequence $||\xi_n(\tau)||$ is uniformly bounded according
to lemma 5.3. The Lesbegue theorem (see, for example,[28]) implies that
the integral in the right-hand side of formula
\r{5.2} tends to zero. Therefore, $||v_n(t)||
\to_{n\to\infty} 0$, so that
$$
||V_n(t) (\hat{Z}_n^{-1}\hat{H}_n    +    \zeta)^{-1}    P_n\Phi     -
(\hat{Z}_n^{-1} \hat{H}_n +\zeta)^{-1} P_n V(t)\Phi || \to_{n\to \infty} 0.
\tag{5.3}
$$
for $\Phi  =  (\hat{H}+\zeta)^{-1}v$.
Property \r{5.3}  is satisfied for
all $\Phi   \in  D(H)$,  on the dense subset of   $\Cal  B$.
Therefore, property  \r{5.3}  is satisfied for all
$\Phi  \in {\Cal B}$. Furthermore,
$$
\matrix
||V_n(t) ((H+\zeta)^{-1}P_n    -    P_n     (H+\zeta)^{-1})     \Phi||
\to_{n\to\infty} 0,\\
||(\hat{Z}_n^{-1} \hat{H}_n    +     \zeta)^{-1}     P_n     -     P_n
(H+\zeta)^{-1})V(t) \Phi || \to_{n\to\infty} 0.
\endmatrix
\tag{5.4}
$$
Eqs.\r{5.3} and \r{5.4} imply that
$$
||(V_n(t)P_n - P_n V(t))\tilde{\Phi}|| \to_{n\to\infty} 0
\tag{5.5}
$$
under condition
$\tilde{\Phi} = (\hat{H}+\zeta)^{-1} \Phi$. Relation
\r{5.5} is satisfied
on the dense subset $D(\hat{H})$ of
$\Cal B$. Therefore, it is satisfied on $\Cal B$.
First statement  of  theorem  3 is proved.  Second statement is proved
analogously.
\enddemo

This work  was supported by the Russian Foundation for Basic Research,
project 99-01-01198.

\newpage

\head
References
\endhead

1. N.N.Bogoliubov and D.V.Shirkov,
"Introduction to the Theory of Quantized Fields", N.-Y., 1959.

2. F.A.Berezin and L.D.Faddeev,
Remark on the Schrodinger equation with singular potential,
{\it Doklady Akad.Nauk SSSR}
{\bf   137}   (1961),
1011-1014.

3. S.Albeverio, F.Gesztesy, R. Hoegh-Krohn, H.Holden,
"Soluble Models in Quantum Mechanics", Springer-Verlag, 1988.

4. Yu.G.Shondin, Quantum mechanical models in ${\Bbb  R}^n$ which are
associated with an extension of the energy operator in the Pontriagin
space, {\it  Teoreticheskaya i Matematicheskaya Fizika} {\bf  74}  (1988),
331-344.

5. Yu.M.Shirokov, Strongly singular potentials in one-dimensional
quantum mechanics,
{\it  Teoreticheskaya i Matematicheskaya Fizika}
{\bf  41}  (1979), 291-302.

6. Yu.M.Shirokov, Strongly singular potentials in three-dimensional
quantum mechanics,
{\it  Teoreticheskaya i Matematicheskaya Fizika}
{\bf  42}  (1980) 45-49.

7. Yu.G.Shondin, Singular point perturbations of  odd operator in
${\Bbb Z}_2$-graduated space,
{\it Matematicheskie Zametki}
{\bf 66} (1999), 924-940.

8. M.I.Neiman-zade and A.A.Shkalikov, Shrodinger operators with singular
potentials from the multiplicator spaces,
{\it Matematicheskie Zametki} {\bf 66} (1999),  722-733.

9. A.M.Savchuk and A.A.Shkalikov, Sturm-Liouville operators
with singular potentials,
{\it Matematicheskie Zametki} {\bf 66} (1999), 924-940.

10. V.D.Koshmanenko,  Perturbations  of  self-adjoint   operators   by
singular bilinear forms,
{\it Ukrainskii Matematicheskii Zhurnal},
{\bf 41} (1989), 3-19.

11. T.V.Karataeva and V.D.Koshmanenko, Generalized sum of operators,
{\it Matematicheskie Zametki} {\bf 66} (1999), 671-681.

12. V.D.Koshmanenko,  Singular perturbations  with  infinite  coupling
constant,
{\it Funct.  Analis i ego Prilozheniya (Functional  Analysis  and  Its
Applications)}, {\bf 33} (1999), N2,  81-84.

13. A.M.Chebotarev,     Symmetric     form     of    the    stochastic
Hadson-Parthasarathy equation,
{\it Matematicheskie Zametki} {\bf 60} (1996), 726-750.

14. V.G.Danilov, V.P.Maslov and V.M.Shelkovich,
Algebras of  singularities  of  solutions  of  quasilinear strictly
hyperbolic first-order systems,
{\it  Teoreticheskaya i Matematicheskaya Fizika}
{\bf 114} (1998) 3-55.

15. N.I.Ahiezer  and  I.M.Glasman,  "Theory  of  Linear  Operators  in
Hilbert Spaces", Nauka, Moscow, 1966.

16. F.A.Berezin, On a Lee model,
{\it Matematicheskii Sbornik},  {\bf 60} (1963), 425-453.

17. O.I.Zavialov,  Wick  polynomials  in  the indefinite inner product
space, {\it  Teoreticheskaya i Matematicheskaya Fizika},
{\bf 16} (1973), 145-156.

18. I.S.Iokhvidov,   M.G.Krein,   Spectral   Theory  of  Operators  in
Indefinite Inner Product Spaces,
{\it Trudy Moskovskogo Matematicheskogo Obshestva},  {\bf  5}  (1956),
367-432.

19. M.A.Naimark,  An analog of the Stone theorem for indefinite  inner
product space,
{\it Doklady Akad.Nauk SSSR},
{\bf 170} (1966) 1259-1261.

20. Shah  Tao-Shing,  {\it  On conditionally positive-definite
generalized functions},  Scientia  sinica,  {\bf  11}  (1962)
1147-1168.

21. V.P.Maslov and O.Yu.Shevdov,  On the axiomatics of  quantum  field
theory with ultraviolet cutoff,
{\it Matematicheskie Zametki}, {\bf 63} (1998) 147-150.

22. T.Kato,  "Perturbation Theory for Linear Operators",  Springer-Verlag,
1966.

23. E.F.Trotter, "Approximation  of  semi-groups of operators"
{\it Pacific J. of Math.} {\bf 8} (1958) 887-919.

24. O.Yu.Shvedov, On Maslov canonical operator in abstract spaces,
{\it Matematicheskie Zametki}, {\bf 65} (1999) 437-456.

25. O.Yu.Shvedov, On Maslov complex germ in abstract spaces,
{\it Matematicheskii Sbornik}, {\bf 190} (1999), N10, 123-157.

26. L.V.Kantorovich and G.P.Akilov, "Functional Analysis",
Nauka, Moscow, 1984.

27. S.G.Krein, "Linear Differential Equations in Banach Space",
Nauka, Moscow, 1967.

28. A.N.Kolmogorov and S.V.Fomin,  "Elements of Functions  Theory  and
Functional Analysis", Nauka, Moscow, 1989.

29. L.S.Pontriagin,  Hermitian operators in indefinite  inner  product
spaces.
{\it Izv. An. SSSR, ser. matemat.}, {\bf 8} (1944), N6,
243-280.

\enddocument